\documentclass[final,5p,times,twocolumn]{elsarticle}

\usepackage[cmex10]{amsmath}
\usepackage{array}
\usepackage{url}

\usepackage{graphicx}
\graphicspath{{./}}
\DeclareGraphicsExtensions{.png}
\usepackage{subfig}

\usepackage{amssymb}
\usepackage{amsthm}

\usepackage{algorithmic}
\usepackage{algorithm}

\usepackage[T1]{fontenc}
\usepackage{textcomp}
\usepackage{lmodern}

\par
\begin{document}
\begin{frontmatter}

\title{Automatic Performance Debugging of SPMD-style Parallel Programs}

 \author[label1,label4]{Xu Liu}
\ead{xu.liu@rice.edu}

\author[label1]{Jianfeng Zhan\corref{cor1} \fnref{fn1}}
\cortext[cor1]{Corresponding author}
\ead{jfzhan@ncic.ac.cn}
\fntext[fn1]{Tel:010-62601006;}

\author[label3]{Kunlin Zhan}
\ead{zhankunlin@ncic.ac.cn}

\author[label2]{Weisong Shi}
\ead{weisong@wayne.edu}

\author[label1,label3]{Lin Yuan}
\ead{yuanlin@ncic.ac.cn}

\author[label1]{Dan Meng}
\ead {md@ncic.ac.cn}

\author[label1]{Lei Wang}
\ead {wl@ncic.ac.cn}

\address[label1]{Institute of Computing Technology, China Academy of Sciences, Beijing 100190, China}
\address[label2]{Department of Computer Science, Wayne State University}
\address[label3] {Graduate University of Chinese Academy of Sciences}
\address[label4] {Department of Computer Science, Rice University}

\begin{abstract}
Automatic performance debugging of parallel applications includes two main steps: locating performance bottlenecks and uncovering their root causes for performance optimization. Previous work fails to resolve this challenging issue in two ways: first, several previous efforts automate locating bottlenecks, but present results in a confined way that only identifies performance problems with apriori knowledge; second, several tools take exploratory or confirmatory data analysis to automatically discover relevant performance data relationships, but these efforts do not focus on locating performance bottlenecks or uncovering their root causes.

The simple program and multiple data (SPMD) programming model is widely used for both high performance computing and Cloud computing. In this paper, we design and implement an innovative system, \emph{AutoAnalyzer}, that automates the process of debugging performance problems of SPMD-style parallel programs, including data collection, performance behavior analysis, locating bottlenecks, and uncovering their root causes. AutoAnalyzer is unique in terms of two features: first, \emph{without any apriori knowledge}, it automatically locates bottlenecks and uncovers their root causes for performance optimization; second, it is lightweight in terms of the size of performance data to be collected and analyzed. Our contributions are three-fold: first, we propose two effective clustering algorithms to investigate the existence of performance bottlenecks that cause process behavior dissimilarity or code region behavior disparity, respectively; meanwhile, we  present two searching algorithms to locate bottlenecks; second, on a basis of the rough set theory, we propose an innovative approach to automatically uncovering root causes of bottlenecks; third, on the cluster systems with two different configurations, we use two production applications, written in Fortran 77, and one open source code---MPIBZIP2 (\url{http://compression.ca/mpibzip2/}), written in C++, to verify the effectiveness and correctness of our methods. For three applications, we also propose an experimental approach to investigating the effects of different metrics on locating bottlenecks.

\end{abstract}

\begin{keyword}
SPMD parallel programs \sep automatic performance debugging \sep performance bottleneck \sep root cause analysis \sep performance optimization


\end{keyword}

\end{frontmatter}


\section{Introduction}
How to improve the efficiency of parallel programs is a challenging issue for programmers, especially non-experts without the deep knowledge of computer science, and hence it is a crucial task to develop an automatic performance debugging tool to help application programmers analyze parallel programs' behavior, locate performance bottlenecks (in short \emph{bottlenecks}), and uncover their root causes for performance optimization.

Although several existing tools can automate analysis processes to some extent, previous work fails to resolve this issue in three ways.
First, with traditional performance debugging tools \cite{ref_14} \cite{hpctookit} \cite{ref_36}, though data collection processes are often automated,
detecting bottlenecks and uncovering their root causes need great manual efforts.
Second, several previous efforts can only automatically identify critical bottlenecks \emph{with apriori knowledge} specified in terms of either \emph{the execution patterns} that represent situations of inefficient behaviors \cite{ref_2} \cite{ref_3} \cite{ref_4} \cite{ref_5} or \emph{the predefined performance hypotheses/thresholds} \cite{ref_7} \cite{ref_9} or \emph{the decision tree classification} trained by microbenchmarks \cite{ref_28}.
Third, while a lots of existing tools \cite{ref_16} \cite{ref_22} \cite{ref_23} \cite{ref_24} \cite{ref_27} \cite{ref_28} \cite{ref_29} \cite{ref_35} \cite{ref_16} take exploratory or confirmatory data analysis approaches to automatically discovering relationships of relevant performance data, these efforts do not focus on locating performance bottleneck and uncovering their root causes of performance bottlenecks.

The SPMD \cite{SPMD} programming model is widely used for high performance computing \cite{INTP}. Recently,  as an instance \cite{Granules}, Mapreduce-like techniques \cite{Mapreduce} \cite{Transformer} also promote the wide use of the SPMD programming model in Cloud computing \cite{INTP} \cite{SC-MTC} \cite{DawningCloud}. This paper focuses on how to automate the process of debugging performance problems of SPMD style programs: including collecting performance data, analyzing application behavior, detecting bottlenecks, and uncovering their root causes, \emph{but not including performance optimization}. To that end, we design and implement an innovative system, \emph{AutoAnalyzer}.

Without human involvement, our tool uses source-to-source transformation to automatically insert \emph{the instrumentation code} into the source code of a parallel program, and divide the whole program into \emph{code regions, each of which is a section of code executed from start to finish with one entry and one exit}. For a SPMD-style parallel program, if we exclude code regions in the master process responsible for the management routines, each process or thread should have similar behavior. At the same time, if a code region takes up a trivial proportion of a program's running time, the performance improvement of the code region will contribute little to the overall performance of the program. From the above intuition, in this paper we pay attentions to two types of performance bottlenecks: bottlenecks that cause process or thread behavior dissimilarity, which we call \emph{dissimilarity bottlenecks},  and bottlenecks that cause code region behavior disparity---significantly different contributions of code regions to the overall performance, which we all \emph{disparity bottlenecks}.  After collecting the performance data of code regions from four hierarchies: application, parallel interface, operating system, and hardware, AutoAnalyzer proposes a series of innovative approaches to searching code regions that are dissimilarity and disparity bottlenecks and uncovering their root causes for performance optimization.
Our contributions are concluded as follows:
 	
 \begin{itemize}
 \item For SPMD-style parallel applications,  we utilize two effective clustering algorithms to investigate the existence of performance bottlenecks that cause process behavior dissimilarity or code region behavior disparity, respectively; if there are bottlenecks, we present two searching algorithms to locate performance bottlenecks.
 \item On a basis of the rough set theory, we propose an innovative approach to automatically uncovering root causes of bottlenecks.
 \item We design and implement AutoAnalyzer. On the cluster systems with two different configurations, we use two production applications and one open source code---MPIBZIP2 to verify the effectiveness and correctness of our system.
 We also investigate the effects of different metrics on locating bottlenecks.
 our experiment results showed for three applications, our proposed metrics outperforms the cycles per instruction (CPI) and the wall clock time in terms of locating disparity bottlenecks.
 \end{itemize}

The rest of this paper is organized as follows: Section \ref{problem} formulates the problem. Section \ref{related} outlines the related work, followed by the description of our solution in Section \ref{solution}.  The implementation and evaluation of AutoAnalyzer are depicted in Section \ref{implementation} and Section \ref{evaluation}, respectively. Finally, concluding remarks are listed in Section \ref{conclusion}.

\section{Problem Statement} \label{problem}

\begin{figure}[h]
\centering
\includegraphics[width=2.5in]{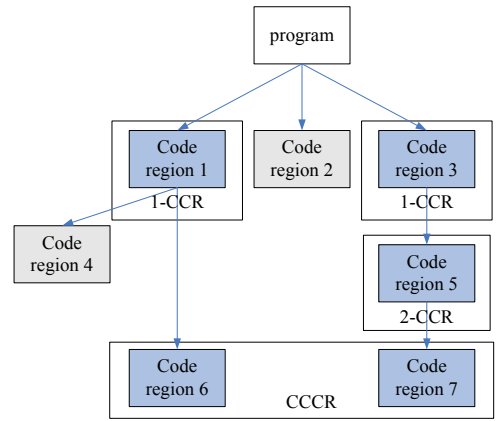}
\caption{The code region tree of a parallel program.}\label{code_region_tree}
\end{figure}

\emph{A code region is a section of code that is executed from start to finish with one entry and one exit}. A code region can be a function, subroutine or loop, which can be nested within another one. After dividing the whole program into $n$ code regions $CR_{j, \; j=1...n}$, we organize $CR_{j, \; j=1 \cdots n}$ as a tree structure with the whole program as the root. According to the definition of the tree structure, for any node $CR_{j}$, its depth is the length of the path from the root to $CR_{j}$. For example, in Fig.\ref{code_region_tree}, the depth of \emph{code region $1$} is one. We call a code region of the depth $L$ an $L$-code region.

In our system, to accurately measure the contribution of each code region to the overall performance of the program, we require
that \emph{code regions that have the same depth can not be overlapped}. For code regions with different depths, we encourage the nesting of code regions because deep nesting leads to fine granularity,
which is helpful in narrowing the scope of the source code in locating bottlenecks.
For example, in Fig.\ref{code_region_tree}, for two 1-code regions, \emph{code region $1$} and \emph{code region $2$} do not intersect. For \emph{code region $1$}, its two children nodes: \emph{code region $4$} and \emph{code region $6$} are nested within it.

For a parallel program, if its processes or threads have similar behavior,  performance vectors of all processes or threads should be classified into one cluster, or else \emph{there are dissimilarity bottlenecks, indicating load imbalance}.  For each code region, if we average its performance data among all processes or threads, we can measure its contribution to the overall performance. We will identify a code region that takes up a significant proportion of a program's running time and has the potential for performance improvement as a disparity bottleneck.
Of course, we can not exhaust all types of bottlenecks, since users hope programs to run faster and faster.

Our work focuses on how to automatically locate dissimilarity and disparity bottlenecks, and uncover their root causes for performance optimization. However,  automatic performance optimization is not our target.



\section{Related Work} \label{related}

\begin{table}[h]
\caption{The comparison of different systems. Yes indicates it is automatic; else not.} \label{comparision}

\begin{minipage}{2.6in}
\begin{tabular}{|p{0.58in}|p{0.38in}|p{0.45in}|p{0.32in}|p{0.35in}|p{0.42in}|}
\hline
\centering
system &data collection &behavior analysis &bottle necks &root causes &optimi zation\\\hline
HPC Viewer & yes & no & no & no & no\\\hline
HPC TOOLKIT   & yes & no & no & no & no \\\hline
TAU  & yes & no & no & no & no \\\hline
EXPERT & yes & yes & yes\footnote{with apriori knowledge} & no & no \\\hline
Paradyn  & yes & yes & yes\footnote{with apriori knowledge} & no & no \\\hline
Aksum & yes & yes & yes\footnote{with apriori knowledge} & no & no \\\hline
Perf Explorer   & yes &yes & no & no & no \\\hline
Tallent {\em et al.} \cite{ref_17}  & /  &/ &/ & / & yes \\\hline
Auto Analyzer   & yes &yes & yes & yes & no \\\hline
\end{tabular}
\end{minipage}
\end{table}

Table \ref{comparision} summarizes the differences of the related systems from five perspectives: data collection, behavior analysis, bottleneck detection, uncovering root causes and performance optimization. Hollingsworth et al \cite{ref_33} proposes a plan to develop a test suite for verifying the effectiveness of different tools in terms of locating performance bottlenecks. If it succeeds, this test suite can provide a benchmark for evaluating the accuracy of locating bottlenecks for different tools in terms of the false positive and the false negative. Unfortunately, this project seems ended without updating its web site.


The traditional approach for performance debugging is through automated data collection and visualizing performance data, while performance analysis and code optimization need great manual efforts. With this approach, application programmers need to learn  appropriate tools, and rely on their expertise to interpret data and its relation to the code \cite{ref_27} so as to optimize the code. For example, HPCViewer \cite{ref_14}, HPCTOOLKIT \cite{hpctookit}, and TAU \cite{ref_36} display the performance metrics through a graphical user interface. Users depend on their expertise to choose valuable data, which is hard and tedious.

With apriori knowledge, previous work proposes several automatic analysis solutions to identify critical bottlenecks.
The EXPERT system \cite{ref_2} \cite{ref_3} \cite{ref_4} \cite{ref_5} describes performance problems using a high level of abstraction in terms of execution patterns that result from an inefficient use of the underlying programming models,
and performs trace data analysis using an automated pattern-matching approach \cite{ref_5}. The Paradyn parallel performance tool \cite{ref_7} starts searching for bottlenecks by issuing instrumentation requests to collect data of a set of pre-defined performance hypotheses for the whole program. Paradyn starts its search by comparing the collected performance data with the predefined thresholds, and the instances where the measured value for the hypothesis exceeds the threshold are defined as bottlenecks \cite{ref_9}. Paradyn starts a hierarchical search of the bottlenecks, and refines this search by using stack sampling \cite{ref_11} and pruning the search space through considering the behavior of the application during previous runs \cite{ref_8}. Using a decision tree classification, which is trained by the microbenchmarks that demonstrate both efficient and inefficient communication, Vetter et al \cite{ref_28} automatically classify individual communication operations, and reveal the cause of communication inefficiencies in the application. The Aksum tool \cite{ref_34} automatically performs multiple runs of a parallel application and detects performance bottlenecks by comparing the performance achieved varying the problem size and the number of allocated processors.  The key idea in the work of \cite{ref_38} \cite{ref_39} is to extract performance knowledge from parallel design patterns or model that represent structural and communication patterns of a program for performance diagnosis.

Several previous efforts propose exploratory or confirmatory data analysis  \cite{ref_27} or fuzzy set method \cite{ref_31} to automated discoveries of relevant performance data.
The PerfExplorer tool \cite{ref_22} \cite{ref_23} \cite{ref_24} addresses the need to manage large-scale data complexity using techniques such as clustering and dimensionality reduction, and performs automated discovery of relevant data relationships using comparative and correlation analysis techniques. By clustering thread performance for different metrics, PerfExplorer should discover these relationships and which metrics best distinguish their differences. Calzarossa {\em et al.} \cite{ref_16} proposes a top-down methodology towards automatic performance analysis of parallel applications: first, they focuses on the overall behavior of the application in terms of its activities, and then they consider individual code regions and activities performed within each code region. Calzarossa {\em et al.} \cite{ref_16} utilizes clustering techniques to summarize and interpret the performance information by identifying patterns or groups of code regions characterized by a similar behavior. Ahn {\em et al.} \cite{ref_29} use several multivariate statistical analysis techniques to analyze parallel performance behavior, including cluster analysis and F-ratio, factor analysis, and principal component analysis. Ahn {\em et al.} \cite{ref_29} show how hardware counters could be used to analyze the performance of multiprocessor parallel machines. The primary goal of the SimPoint system \cite{ref_32} is to reduce long-running applications down to tractable simulations. Sherwood {\em et al.} \cite{ref_32} define the concept of basic block vectors, and use those concepts to define the behavior of blocks of execution, usually one million instructions at a time. Truong {\em et al.} \cite{ref_30} \cite{ref_31} propose a fuzzy set approach to search bottlenecks. However, it does not intend to uncover the root causes of bottlenecks. Tallent {\em et al.} \cite{ref_17} propose the approaches to measure and attribute parallel idleness and parallel overhead of multi-threaded parallel applications.

Tiwari {\em et al.} \cite{ref_35} describes a scalable and general-purpose framework for auto-tuning compiler-generated code, which generates in parallel a set of alternative implementations of computation kernels and automatically selects the one with the best-performing implementation. Tu {\em et al.} \cite{JSC} \cite{tu_pdp} propose a new parallel computation model to characterize the performance effects of the  memory hierarchy on multi-core clusters in both vertical and horizontal levels. Babu {\em et al.} \cite{debug_mapreduce} make
a case for techniques to automate the setting of tuning parameters for MapReduce programs. Zhang {\em et al.} \cite{preciseTracer} propose a precise request tracing approach to debug performance problems of multi-tier services of black boxes.

Our system has two distinguished differences from other systems as shown in Table \ref{comparision}: first, in addition to automatic performance behavior analysis, we automatically locate bottlenecks of SPMD-style parallel programs without apriori knowledge; second, we automatically uncover the root causes of bottleneck for performance optimization. With regard to proposing performance vectors to represent behavior of parallel application, AutoAnalyzer is similar to the work in \cite{ref_16} \cite{ref_22} \cite{ref_23} \cite{ref_24} \cite{ref_30}, but we investigate the effect of different metrics on locating bottlenecks. Different from PerfExplorer \cite{ref_22} \cite{ref_23} \cite{ref_24}, which leverages sophisticated clustering techniques, AutoAnalyzer adopts comparatively simple clustering algorithms, which are lightweight in terms of the size of performance data to be collected and analyzed.

\section{Our Solution} \label{solution}
This section includes four parts: Section \ref{overview} summarizes our approach, followed by the description of the approaches to investigating the existence of bottlenecks in Section \ref{existence}. How to locate bottlenecks is given out in Section \ref{bottlenecks}. Finally, we propose an approach to uncovering the root causes of bottlenecks.

\subsection{Summary of our approach} \label{overview}
Our method includes four major steps: instrumentation, collecting performance data, locating bottlenecks, and uncovering their root causes.

First, we instrument a whole parallel program into code regions. Our tool uses source-to-source transformation to automatically insert instrumentation code into the source code, which requires no human involvement.

Second, we collect performance data of code regions. For each process or thread, we collect the following performance data of code regions: (1) application-level performance data: \emph{wall clock time} and \emph{CPU clock time}; (2) hardware counter performance data: \emph{clock cycle, instructions retired, L1 cache miss, L2 cache miss, L1 cache access, L2 cache access}; (3) communication performance data:  \emph{MPI communication time}---the executing time in MPI library and \emph{MPI communication quantity}---the quantity of data transferred by the MPI library; (4) operation system level performance data: \emph{disk I/O quantity}---the quantity of data read and written by disk I/O. On a basis of hardware counter performance data, we obtain two derived metrics: \emph{L1 cache miss rate} and \emph{L2 cache miss rate}. For example L1 cache miss rate can be obtained according to the formula---\emph{((L1 cache miss) / (L1 cache access))}.

Third, we utilize two clustering approaches to investigating the existence of bottlenecks. If there are bottlenecks, we use two searching algorithms to locate bottlenecks.

Finally, on a basis of the rough set theory, we present an approach to uncovering the root causes of bottlenecks.

\subsection {Investigating the existence of bottlenecks} \label{existence}

In this section, we present how to investigate existence of dissimilarity bottlenecks and disparity bottlenecks, respectively.

\subsubsection{The existence of dissimilarity bottlenecks}\label{existenceDissimilarity}
For a SPMD program, each process or thread is composed of the same code regions. If we exclude code regions in the master process responsible for the management routines, the high behavior similarity of each process or thread indicates the balance of workload dispatching and resources utilizing, and vice versa \cite{ref_16}. So we use a similarity analysis approach to investigate the existence of dissimilarity bottlenecks.

The performance similarity is analyzed among all participating processes or threads to discover the discrepancy. We presume that the whole program is divided into $n$ code regions, and the whole program has $m$ processes or threads.
In our approach, each process\textquotesingle{ } or thread\textquotesingle{ } performance is represented by a vector $\overrightarrow{V_{i}}$, where $i$ is the process or thread rank. $T_{it}$ represents the performance measurement of the $t_{th}$ code region in the $i_{th}$ process or thread. So $\overrightarrow{V_{i}}$ is described as $\overrightarrow{V_{i}}=\left(T_{i1},T_{i2}\cdots,T_{in}\right)$.

We define the Euclidean distance---$Dist_{ij}$ of two vectors $\overrightarrow{V_{i}}$ and $\overrightarrow{V_{j}}$ in Equation(\ref{dist}).

\begin{equation}\label{dist}
Dist_{ij}=\sqrt{\left(T_{i1}-T_{j1}\right)^{2}+\cdots+\left(T_{in}-T_{jn}\right)^{2}}
\end{equation}

We choose the CPU clock time of each code region as the main measurement. Different from the wall clock time, the CPU clock time only measures  the time during which the processor is actively working on a certain task, while the wall clock time measures the total time for a process to complete. We also observe the effect of choosing different metrics---the wall clock time on locating dissimilarity bottlenecks in Section \ref{different_metrics}

On a basis of Equation(\ref{dist}), we present a simplified OPTICS clustering method \cite{ref_1}---Algorithm \ref{correlation_algorithm} to classify all processes or threads.
We choose the simplified OPTICS clustering method because it has advantage in discovering isolated points.
In this approach, the performance vector of each process or thread is considered as a point in an n-dimension space.
A set of points is classified into one cluster if the point density in the area, where these point scattered, is larger than the defined threshold.
If a point is not included into any clusters, we consider it an isolated point, which is also a new cluster.


\begin{algorithm}
\caption{The simplified OPTICS clustering algorithm \textbf{\{\}}} \label{correlation_algorithm}
\algsetup{
linenosize=\small,
linenodelimiter=.
}
\begin{algorithmic}[1]
\REPEAT
\STATE select a performance vector $\overrightarrow{V_{p}}$ not belonging to any clusters.
\STATE count=0;
\FOR   {each point $\overrightarrow{V_{q}}$ ($q \neq p$) in the n-dimension space}
\IF    {(distance($\overrightarrow{V_{p}}$ , $\overrightarrow{V_{q}}$) < threshold)}
\STATE count++;
\STATE //We set the threshold as 10\% $\times$ length ($\overrightarrow{V_{p}}$).
\ENDIF
\ENDFOR
\IF   {$count > count\_threshold$}
\STATE confirm that this is a new cluster.
\ENDIF
\UNTIL {all vectors are compared.}

\end{algorithmic}
\end{algorithm}


For a SPMD program, if Algorithm \ref{correlation_algorithm} classifies performance vectors of all processes or threads into one cluster, indicating all processes have similar performance behavior, we confirm that there are no dissimilarity bottlenecks, or else there are dissimilarity bottlenecks.



\subsubsection{The existence of disparity bottlenecks} \label{existenceDisparity}

For each code region, if we average performance data among all processes or threads, we can measure its contribution to overall performance. We will identify a code region that takes up a significant proportion of a program's running time and has the potential for performance improvement as a disparity bottleneck.

We propose a single normalized metric, named \emph{the code region normalized metric (in short, CRNM)}, as the measurement basis for performance contribution of each code region to the overall performance of the application. For each code region, CRNM is defined in Equation (\ref{CRN}):

\begin{equation} \label{CRN}
CRNM=\frac{CRWT}{WPWT} \star CPI
\end{equation}

In Equation (\ref{CRN}), CRWT is \emph{the wall clock time of the code region}; WPWT is \emph{the wall clock time of the whole program}; CPI is \emph{the average cycles per instruction of each code region}. In Section \ref{different_metrics}, we also investigate the effects of choosing other metrics, e.g., CPI and wall clock time of each code region, on locating disparity bottlenecks.

\begin{figure}[h]
\centering
\includegraphics[width=2.5in]{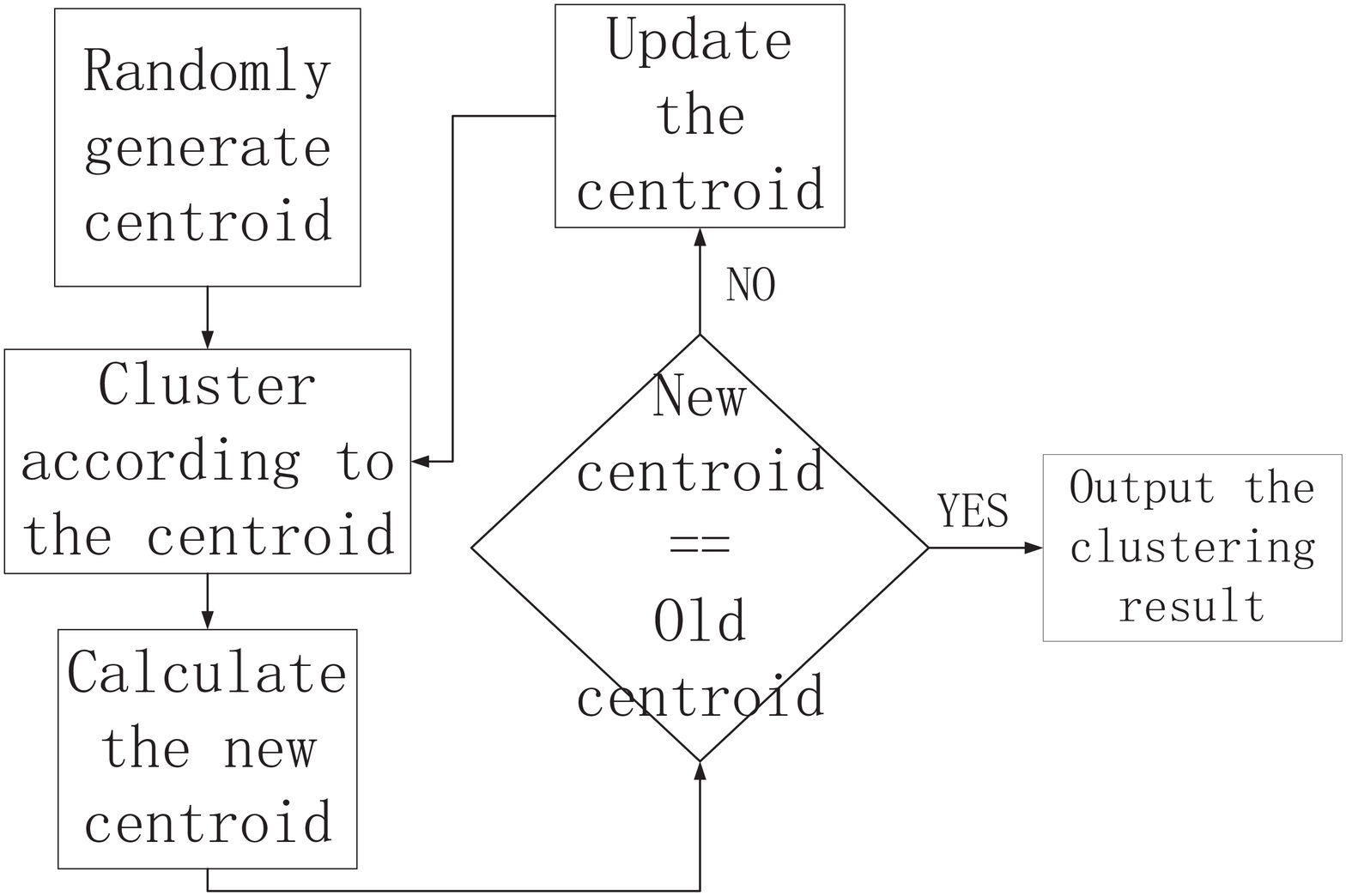}
\caption{The k-means clustering approach \cite{ref_12}.} \label{procedure}
\end{figure}

As shown in Fig.\ref{procedure}, the procedure of searching disparity bottlenecks is as follows:

First, for each processes or thread, we obtain the CRNM value of each code region. If a code region is not on the call path in a process or thread, its CRNM value is zero. Since a SPMD program can contain 'if' statements, we obtain the average value of each code region among all processes or threads.

Second, we use a k-means clustering method \cite{ref_12} to classify each code region according to the average CRNM value. We choose the k-means clustering method because it can classify data into $k$ clusters without user providing the threshold value. We define five severity categories: \emph{very high (4), high (3), medium (2), low (1), and very low (0)}. The k-means clustering method finally classifies each code region into one of the severity categories according to its CRNM value.

Third, if a code region is classified into one of severity categories of \emph{very high} or \emph{high}, we consider it as a critical code region (CCR).

\subsection{ Locating bottlenecks} \label{bottlenecks}

When users confirm there are bottlenecks, they need to locate bottlenecks. We call the code region that is a bottleneck \emph{a critical code region (in short CCR)}. A CCR of the depth $L$ is called an \emph{L-CCR}. If a CCR satisfies the following conditions, we call it \emph{a core of critical code regions (in short, CCCR)}: (1) the CCR is a leaf node in the code region tree; (2) for a CCR, its children nodes are not CCR. For example, in Fig.\ref{code_region_tree}, both code region 6 and code region 7 are CCCR.

We propose a top-down searching algorithm---Algorithm \ref{searching_algorithm} to locate dissimilarity bottlenecks as follows:

\begin{algorithm}
\caption{The searching algorithm for dissimilarity bottlenecks\textbf{\{\}}\\
$n$: the number of code regions;\\
$r$: the number of 1-code region;\\
$m$: the number of process or threads;}
\label{searching_algorithm}
\algsetup{
linenosize=\small,
linenodelimiter=.
}
\begin{algorithmic}[1]
\STATE CCR\_set=null;
\STATE CCCR\_set=null;

\FOR   {each code region $j$, $j=1...n$}
\STATE $T\_backup_{ij} =T_{ij}, i=1...m$.

  \IF    {its depth is greater than one}
  \STATE $T_{ij} =0, i=1...m$.
  \ENDIF
\ENDFOR

\STATE Obtain the clustering results.

\FOR   {each code region $j$, $j=1...n$}
  \IF    {its depth is equal with one}
  \STATE $T_{ij} =0, i=1...m$.
  \STATE Obtain the new clustering results.

    \IF    {the clustering result changes}
    \STATE {Add code region $j$ into CCR\_set.}
    \STATE {Recursively analyze children of code region $j$.}

      \FOR   {each child code region k}
      \STATE $T_{ik}=T\_backup_{ik}, i=1...m$.
      \STATE Obtain the new clustering results.

          \IF    {the clustering result does not change}
          \STATE {Add code region $k$ into CCR\_set.}

              \IF    { (CCR $k$ is a leaf node) or (its any child is not a CCR)}
              \STATE  CCR $k$ is a CCCR.
              \ENDIF

          \ENDIF

      \ENDFOR
    \STATE $T_{ij} =T\_backup_{ij}, i=1...m$.
    \ENDIF
  \ENDIF
\ENDFOR

\IF   {CCR\_set is null}
\STATE {Combine $s$ adjacent 1-code regions into composite code regions without overlapping, $s\geq2$.}
\STATE {Repeat the above analysis.}
  \IF     {CCR\_set is null and $s<(r-1)$}
  \STATE  {increment $s$ and repeat the above analysis.}
  \ENDIF
\ENDIF
\end{algorithmic}
\end{algorithm}

According to Line 17-26 in Algorithm \ref{searching_algorithm}, a CCCR has higher effect on the clustering results than the other children of its parent CCR, and hence we only consider CCCR as dissimilarity bottlenecks, on which users should focus for performance optimization. If the number of clusters or members of a cluster
change, we think the clustering result changes, or else not.

We also propose a simple searching algorithm to refine the scope of disparity bottlenecks as follows:
\begin{itemize}
  \item If a leaf node $j$ is a CCR, then the code region $j$ is a CCCR.
  \item For a none-leaf CCR $j$, if its severity degree is larger than that of each child node, then we consider the code region $j$ as a CCCR.
\end{itemize}


\subsection {Root Cause Analysis}
In this section, we  introduce the background material of the rough set theory, and present the approaches to recovering roots causes of dissimilarity and disparity bottlenecks, respectively.

\subsubsection{The rough set approach \cite{ref_15} \cite{ref_19}}\label{roughset}

The rough set approach is a data mining method that can be used for classifying vague data. In this paper, we use the rough set approach to uncovering the root causes of dissimilarity and disparity bottlenecks.

We start with introducing some basic terms, including \emph{information system}, \emph{ decision system}, \emph{decision table}, and \emph{core}.

An information system is a pair $\Lambda=(U,A)$, where $U$ is is a non-empty finite set of objects, called the universe, and $A$ is a non-empty finite set of attributes such that $a:U\rightarrow V_{a}$ for every $a\in A$. The set $V_{a}$ is called the value set of $a$.

A decision system is any information system of the form $\Lambda=(U, A\cup {d})$, where $d\notin A$ is the decision attribute. The elements of A are called conditional attributes.

As shown in Table \ref{decision}, a decision table is used to describe the decision system. Each entry of a decision table consists of three parts: \emph{object ID, conditional attributions, and decision attribution}. For example, in Table \ref{decision}, the set of object ID is $\{0, ..., 3\}$, the set of attributions is $\{a_{1}, ..., a_{4}\}$,  and the set of decisions is $\{N, P\}$.

\emph{The core attributions} are the attributions that are critical to distinguishing with the decision attributions. How to find the core attributions is a main research field in the rough set approach. One of the solutions is to create \emph{a discernibility matrix} \cite{ref_15} according to the decision table, and then obtain the core attributions using a discernibility matrix as follows:

For a decision system, its decision-relative discernibility matrix is a symmetric
$n \times n$ with entries $c_{ij}$ given in Equation \ref{cij}. Each entry thus consists of the set of attributions upon which $x_{i}$ and $x_{j}$ differ \cite{ref_15}.


\begin{equation}\label{cij}
  c_{ij, \; i, j=1,...,n} =
  \begin{cases}
   (a \in A | a(x_{i}) \neq a(x_{j}) & \text{if } d(x_{i})\neq d(x_{j}))\\
   (\phi      & \text{otherwise })
  \end{cases}
\end{equation}

A discernibility function $f_{\Lambda}$ for the decision table $\Lambda$ is a Boolean function of $m$ Boolean variables $a_{1}, a_{2},...,a_{m}$ defined in Equation \ref{discernibility}. For example, for Table \ref{decision}, the discernibility functions are shown in Equation \ref{function_example}.

\begin{equation}\label{discernibility}
f_{\Lambda}(a_{1},...a_{m})=\bigwedge \{ \bigvee c_{ij} \mid 1\leq i \leq j \leq n, c_{i,j}\neq \phi \}
\end{equation}

\begin{table}[h]
\centering
\caption{An example of decision table}\label{decision}
\begin{tabular}{|c|c|c|c|c|c|}
\hline

ID & $a_{1}$ & $a_{2}$ & $a_{3}$ & $a_{4}$ & decision\\ \hline
0 & sunny & hot & high & False & N\\ \hline
1 & sunny & hot & high & True & N \\ \hline
2 & overcast & hot & high & False & P\\ \hline
3 & sunny & cool & low & False & P\\ \hline
\end{tabular}
\end{table}

\begin{center}
$\left(\begin{array}{cccc}
\phi & \phi & a_{1} & a_{2}, a_{3}\\
     &\phi & a_{1}, a_{4}  & a_{2}, a_{3}, a_{4}\\
     & & \phi        & \phi \\
     & & &\phi\end{array}\right)$
\par\end{center}

\begin{figure}[h]
\caption{The discernibility matrix for the decision table in Table \ref{decision}.}\label{matrix}
\end{figure}

The core attributions are the same conjunctive terms shared by the discernibility functions of each object, which are defined in Equation \ref{discernibility}.

\begin{equation} \label{function_example}
 \left.\begin{aligned}
        f_{\Lambda}(a_{1},a_{2},a_{3},a_{4})= & (a_{1})\wedge (a_{2} \vee a_{3})\\
                                              & (a_{1} \vee a_{4} )   \wedge ( a_{2} \vee a_{3}  \vee a_{4})
       \end{aligned}
 \right\}
\end{equation}

According to Equation \ref{function_example}, the same conjunctive terms are $\{a_{1}, a_{2}\}$ or $\{a_{1}, a_{3}\}$, which are the core attributions of Table \ref{decision}.



\subsubsection{Root cause analysis}

For performance optimization, users need to know the root causes of bottlenecks. In this section, we propose the rough set theory based approach to uncovering the root causes of dissimilarity and disparity bottlenecks, and give suggestions for performance improvements.

\begin{figure}[h]
\centering
\includegraphics[width=2.5in]{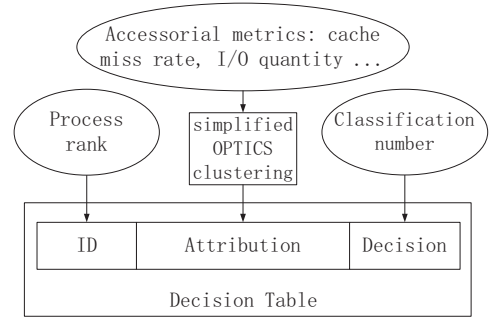}
\caption{The approach to uncovering the root causes of dissimilarity bottlenecks.}\label{approachexternalbottleneck}
\end{figure}

As shown in Fig.\ref{approachexternalbottleneck}, we create the decision table for dissimilarity bottlenecks as follows:
we choose \emph{the rank of each process} as the object ID. We select  \emph{L1 cache miss rate, L2 cache miss rate, disk I/O quantity, network I/O quantity and instructions retired} as five different attributions $a_{k, k=1...5}$.

We take the attribution $a_{1}$ (L1 cache miss rate) as an example. For process $i$, the entry of the decision table corresponding to $a_{1}$ is obtained as follows:

For the performance vector $\overrightarrow{T_{i}}$, where $i=1...m$, we assign $T_{ij}$ with the L1 cache miss rate of \emph{the $j_{th}$ code region} in \emph{process} $i$.

After having created the performance vector, we use the simplified OPTICS clustering algorithm to classify performance data. If $\overrightarrow{T_{i}}$ is classified into \emph{a cluster with the ID of $x$} according to the approach introduced in Section \ref{existenceDissimilarity}, for process $i$, we assign the entry corresponding to $a_{1}$ with $x$.

For \emph{process} $i$, the decision value is the ID of the cluster into which \emph{process} $i$ is classified according to the metrics of \emph{the CPU clock time}.


For disparity bottlenecks, we create the decision table as follows:

We use the code region ID to identify each table entry. We also select L1 cache miss rate, L2 cache miss rate, disk I/O quantity, network I/O quantity and executing instruction number as five different attributions.

We take attribution $a_{1}$ (L1 cache miss rate) as an example. For \emph{code region} $j$, the element of the decision table corresponding to $a_{1}$ is obtained as follows:

For each code region, we obtain the average L1 cache miss rate in all processes or threads. We use the K-means clustering algorithm to classify the average L1 cache miss rates of each code region into five categories: \emph{very high (4) , high (3), medium (2) , low (1), and very low (0)}. For \emph{code region} $j$, if its severity category is higher than \emph{medium}, we assign the entry corresponding to the attribution $a_{1}$ with 1, otherwise 0.

For \emph{code region} $j$, if it is a disparity bottleneck according to the approach proposed in Section \ref{existenceDisparity}, then the decision value is 1, otherwise 0.

After having created the decision table, we obtain the core attributions according to the approaches proposed in Section \ref{roughset}. Since the core attributions are the ones that have dominated effects on the decision, we consider them as the root causes of disparity bottlenecks.

\begin{figure}[h]
\centering
\includegraphics[width=2.5in]{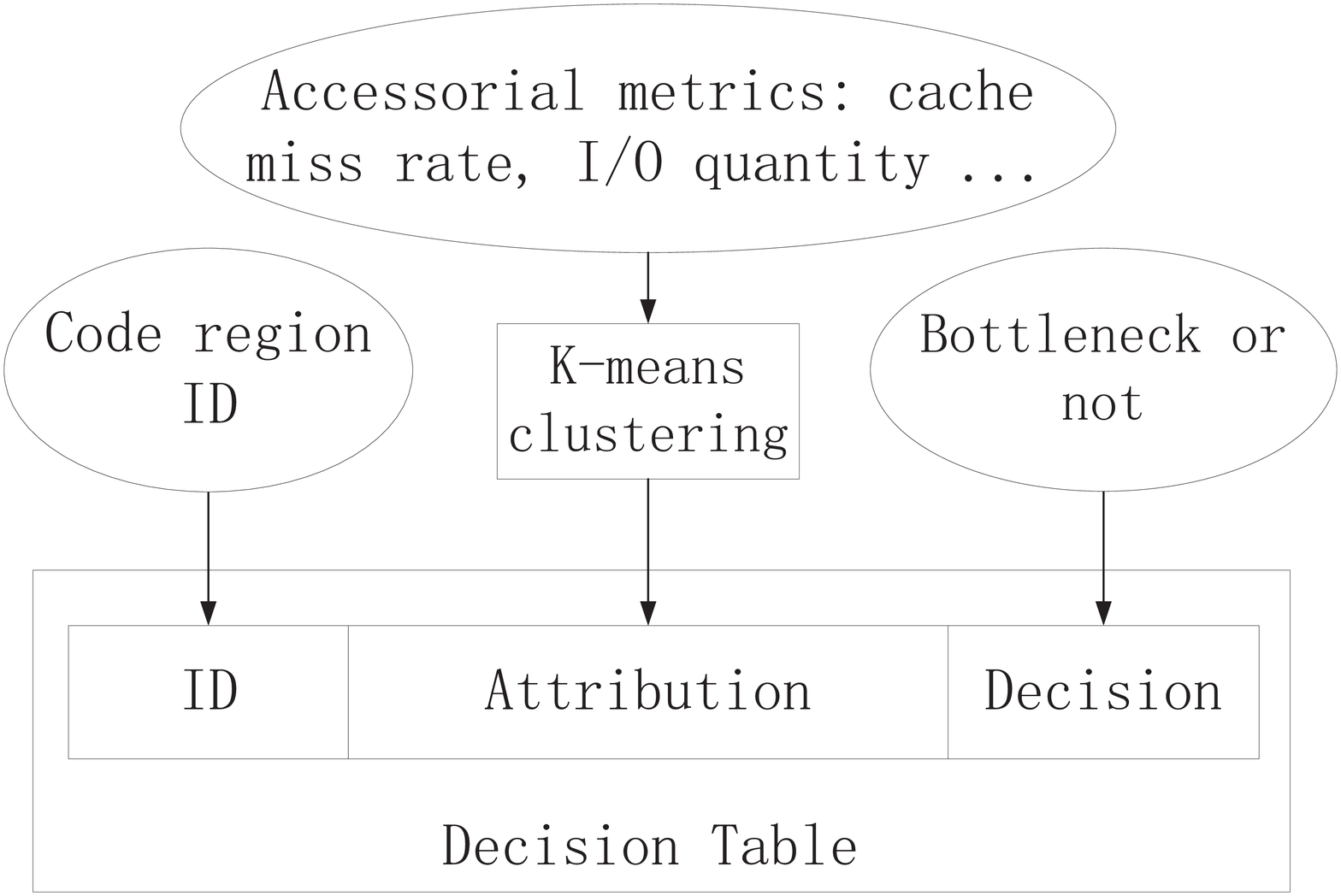}
\caption{The approach to uncovering the root causes of disparity bottlenecks.}\label{approachinternalbottleneck}
\end{figure}

\section{AutoAnalyzer implementation} \label{implementation}
\begin{figure}[h]
\centering
\includegraphics[width=3in]{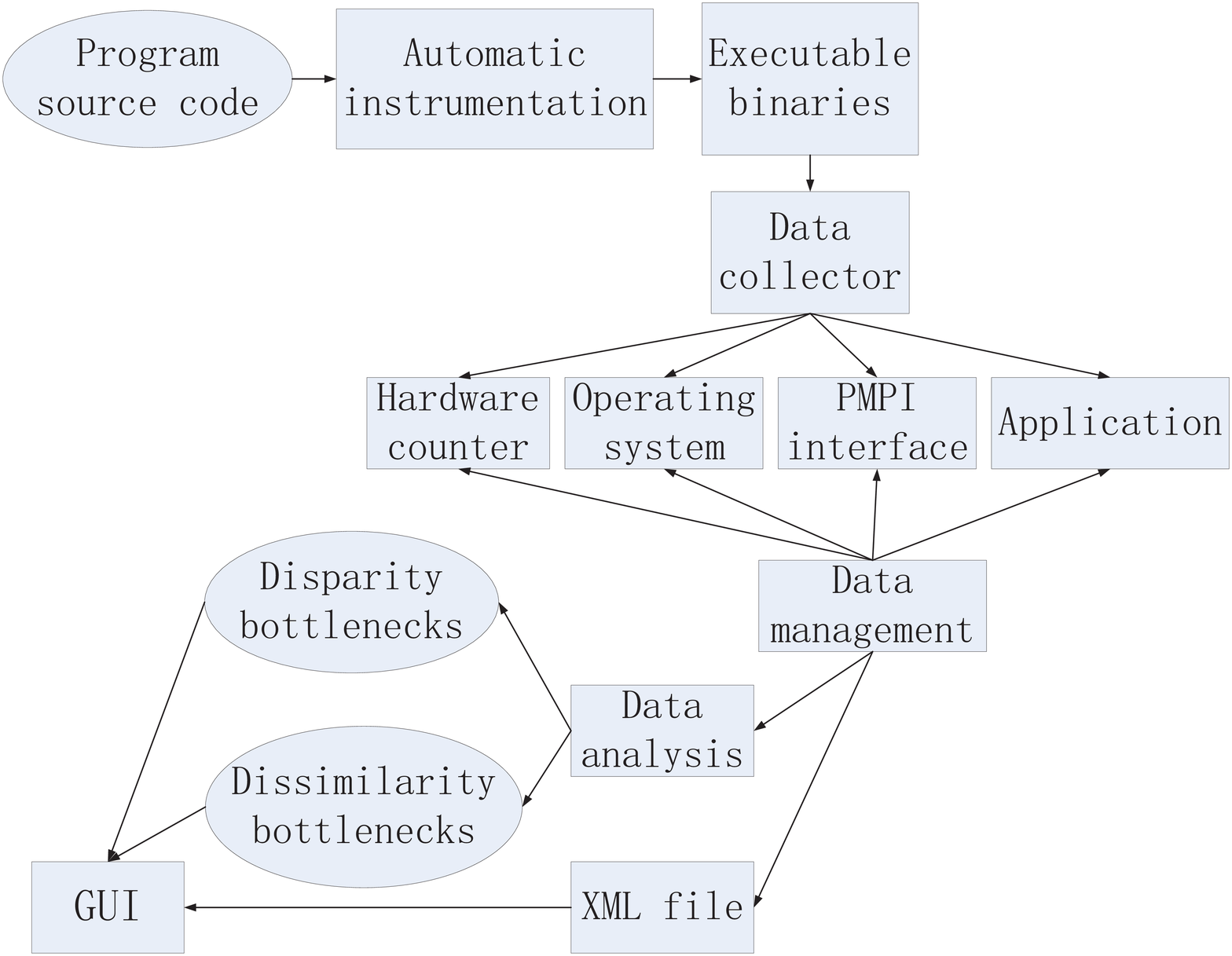}
\caption{The \emph{AutoAnalyzer} Architecture.}\label{architecture}
\end{figure}

In order to evaluate the effectiveness of our proposed methods, we have designed and implemented a prototype, AutoAnalyzer. Presently, AutoAnalyzer supports  debugging of performance problems of SPMD style MPI applications, written in C, C++, FORTRAN 77, and FORTRAN 90. We are also extending our work to MapReduce \cite{Mapreduce} and other data-parallel programming models \cite{Transformer}.  Fig. \ref{architecture} shows AutoAnalyzer architecture.

The major components of \emph{AutoAnalyzer} include \emph{automatic instrumentation, data collector, data management, and data analysis}.

\textbf{Automatic instrumentation}. On a basis of OMPi \cite{ref_17}---a source-to-source compiler, we have implemented the source code level instrumentation. Without human involvement, our tool uses source-to-source transformation to automatically insert instrumentation code. After having parsed the program, the system builds the abstract syntax tree (AST). AST shows program's structure information, e.g., the begin and end of \emph{functions, procedures or loops}. With the structure information, our tool can automatically insert instrumentation codes, and divide a program into code regions.

Our tool supports several instrumentation modes: outer loop, inner loop, mathematical library, parallel interface library like MPI, system call,
C/FORTRAN library, and user-defined functions or procedures. Without any restrictions on instrumentation, a program can be divided into hundreds or thousands of code regions.
For example, after instrumentation, a parallel program of 2,000 lines is divided into more than 300 code regions.
This situation has negative influence on the performance analysis because AutoAnalyzer needs to collect and analyze a large amount of performance data.
To decrease the size of performance data, we propose two solutions: first, we adopt two rounds of analysis. For the first round,  we divide a parallel program into coarse-grained code regions, e.g., per function, for roughly locating bottlenecks; for the second round, we divide \emph{the code regions that are possible bottlenecks} into fine-grained code regions, e.g., loops. Second, users can selectively choose one or more modes to instrument the code, or interact with the GUI of the tool to eliminate, merge, and split code regions.


\textbf{Data collector}. We collect performance data from four hierarchies: application, parallel interface, operating system, and hardware.

In the application hierarchy, we collect the wall clock time and the CPU clock time of each code region. In the parallel interface hierarchy, we have implemented an MPI library wrapper to record MPI routines' behavior of both point-to-point and collective communication. The wrapper is implemented by wrapping the MPI standard profiling interface---PMPI. In the wrapper, we instrumented codes to collect performance data of MPI library, e.g., the executing time and the quantity of data transferred  in MPI library.

In the operating system hierarchy, we use systemtap (\url{http://sourceware.org/systemstap/}) to monitor disk I/O, recording the execution time and quantity of data read and written in I/O operations. Systemtap is based on Kprobe, which is implemented in the Linux kernels. Kprobe can instrument the system calls of the Linux kernel to obtain the executing time and functions' parameters as well as I/O quantity.

In the hardware hierarchy, we use PAPI (\url{http://icl.cs.utk.edu/papi/}) to count hardware events, including L1 cache miss, L1 cache access, L2 cache miss, L2 cache access, and instructions retired.

\textbf{Data management}. We collect all performance data on different nodes and send them to one node for analysis. All data are stored in XML files.

\textbf{Data analysis}. We analyze performance data of code regions so as to search bottlenecks and uncover their root causes. 


Before using AutoAnalyzer, users need to perform the following setup work. Before installing PAPI, they must make sure that the kernel has been patched and recompiled with the PerfCtr or Perfmon patch. Then they can compile the PAPI source code to install it. SystemTap is also dependent upon the installation of several packages: kernel-debuginfo, kernel-debuginfo-common RPMs, and the kernel-devel RPM. Before installing Systemtap, users need to install these packages.  However, with the support of state-of-the-practice operating system deployment tool, like SystemImager, which is open source, we can automate the deployment of AutoAnalyzer.

\section{Evaluation} \label{evaluation}
In this section, we use two production parallel applications, written in Fortran 77, and one open-source parallel application, written in C++, to evaluate the correctness and effectiveness of AutoAnalyzer.

\begin{figure}[h]
\centering
\includegraphics[width=3.5in]{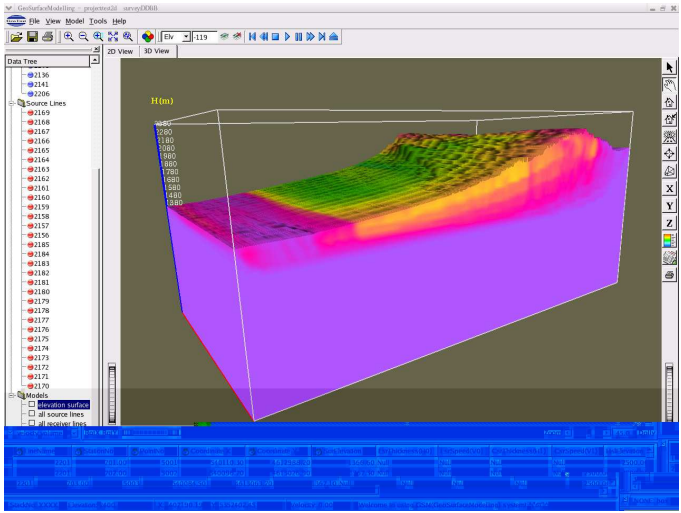}
\caption{The model obtained with \emph{ST}. }\label{ST-GUI}
\end{figure}

The first program is \emph{ST}, which calculates the seismic tomography using a refutations method. ST is on the production use in the largest oil company in China.
Fig.\ref{ST-GUI} shows the model obtained with \emph{ST}.
The second one is a parallel NPAR1WAY module of SAS. SAS is a system widely used in data and statistical analysis. The third one is MPIBZIP2---a parallel implementation of the bzip2 block-sorting file compressor that uses MPI and achieves significant speedup on cluster machines (\url{http://compression.ca/mpibzip2/}).

In Section \ref{ST}, Section \ref{NPAT1WAY}, and Section \ref{MPIBZIP2},  for three applications we choose the CPU clock time as the main performance measurement for searching dissimilarity bottlenecks, and our proposed CRNM as the main performance measurement for disparity bottlenecks, respectively. In Section \ref{different_metrics}, we investigate the effects of different metrics on locating bottlenecks.


\subsection{ST} \label{ST}
In this section, we use a production parallel application of 4307 line codes---ST, to evaluate the effectiveness of our system. To identify a problem, a user of our tools does little to start. The tool automatically instruments the code. After analysis, the tool informs the user about bottlenecks and their root causes. For ST, it took about 2 days for a master student in our lab to locate bottlenecks and rewrite about 200 lines to optimized the code.

\begin{figure}[h]
\centering
\includegraphics[width=2.5in]{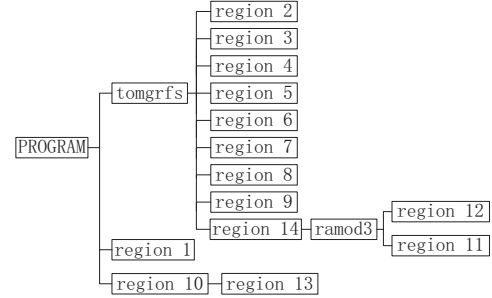}
\caption{The code region tree of ST. \emph{Code region 11, 12} are in subroutine $ramod3$, which is nested in \emph{code region 14}. All code regions contain loops.}\label{STtree}
\end{figure}

Out testbed is a small-scale cluster system, connected with 1000 Mbps networks. Each node has two processors, each of which is AMD Opteron with $64 KB$ L1 data cache, $64 KB$ L1 instruction cache, and $1 MB$ L2 cache. The OS version is $linux-2.6.19$.

In the rest of this section, we give the detail of locating bottleneck and optimizing performance.
Section \ref{ST_B} reports a case study of ST with coarse-grain code regions for locating bottlenecks and optimizing application.
Section \ref{granularity} reports a case study of ST with fine-grain code regions.

\subsubsection{Locating bottlenecks and optimizing the applications}\label{ST_B}

To reduce the number of code regions, AutoAnalyzer support an instrumentation mode that allows a user to select whether to instrument \emph{functions or procedures}, or \emph{outer loops}. In this subsection, we instrument ST into 14 coarse-grain code regions, and Fig. \ref{STtree} shows the code region tree. For ST, a configuration parameter---\emph{the shot number} decides the amount of data input. For this experiment, the shot number is 627.

\begin{figure}[h]
\centering
\includegraphics[width=2.5in]{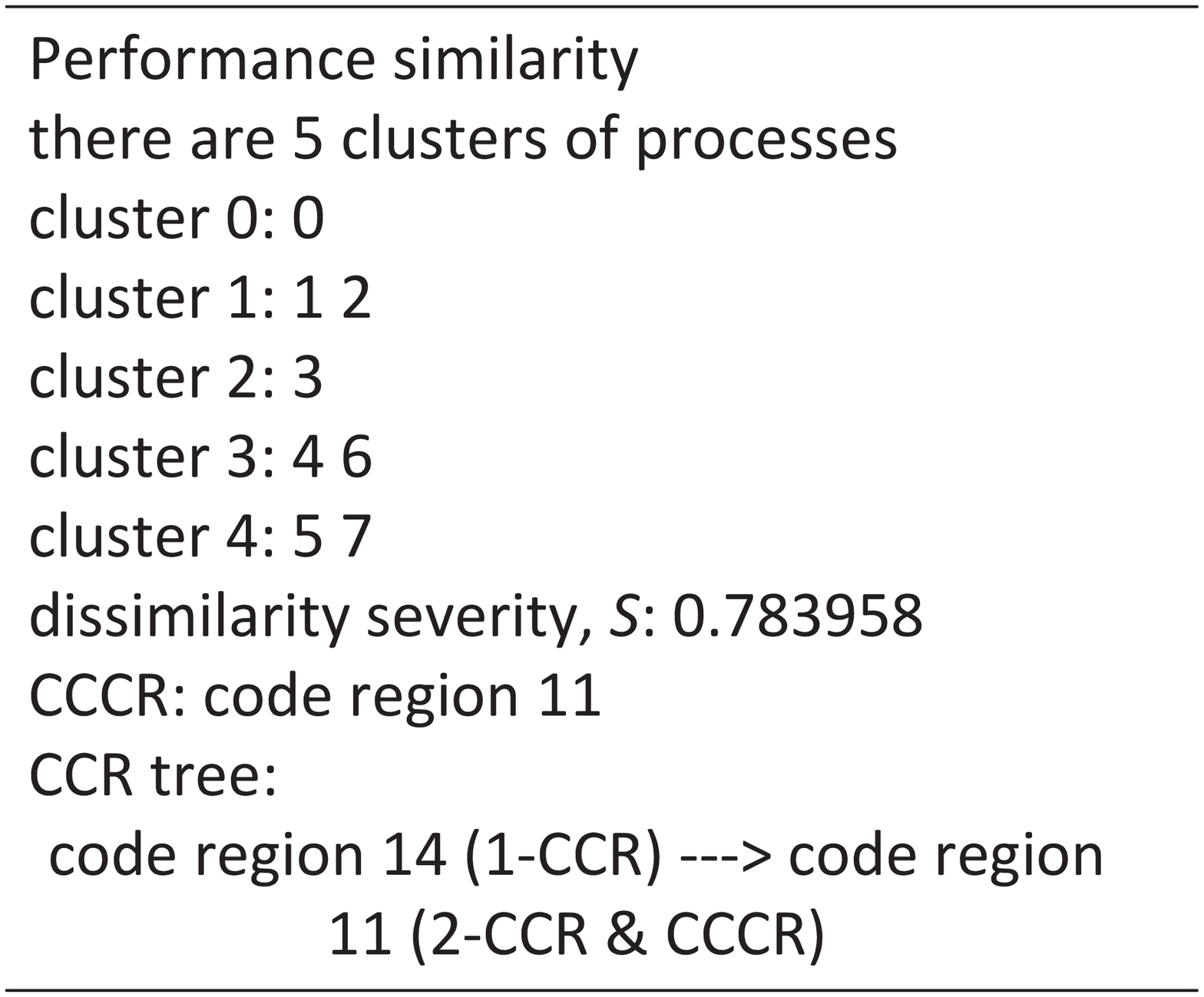}
\caption{The analysis results of similarity measurement.}\label{ST_analysis}
\end{figure}

According to the similarity analysis approach proposed in Section \ref{bottlenecks},  AutoAnalyzer outputs the analysis result for each process behavior of ST, which is shown in Fig.\ref{ST_analysis}.
We can find that all processes are classified into five clusters. 
For a SPMD program, the analysis results indicate that dissimilarity bottlenecks exist. According to the searching result, we can conclude that \emph{code region 11 and code region 14} are CCRs. Since code region 11 is the child node of code region 14, we consider code  region 11 as a CCCR, which is the location of the problem.

We create the decision table to analyze the root causes of \emph{code region 11}.

\begin{table}[h]
\centering
\caption{Decision table for the dissimilarity bottlenecks}\label{externaltable}
\begin{tabular}{|c|c|c|c|c|c|c|}
\hline
ID & $a_{1}$ & $a_{2}$ & $a_{3}$ & $a_{4}$ & $a_{5}$ & D\\ \hline
0 & 0 & 0 & 0 & 0 & 0 & 0\\ \hline
1 & 0 & 0 & 0 & 0 & 1 & 1\\ \hline
2 & 0 & 0 & 0 & 0 & 1 & 1\\ \hline
3 & 1 & 0 & 0 & 0 & 2 & 2\\ \hline
4 & 0 & 1 & 0 & 0 & 3 & 3\\ \hline
5 & 1 & 1 & 0 & 1 & 4 & 4\\ \hline
6 & 1 & 2 & 0 & 1 & 3 & 3\\ \hline
7 & 1 & 2 & 0 & 0 & 4 & 4\\ \hline
\end{tabular}
\end{table}

Table \ref{externaltable} shows the decision table. In the decision table, the attributions $a_{k, k=1,2,3,4,5}$ represents \emph{L1 cache miss rate, L2 cache miss rate, disk I/O quantity, network I/O quantity, and instructions retired}, respectively. Fig.\ref{externalmatrix} shows the discernibility matrix.

\begin{center}
$\left(\begin{array}{p{0.01in}p{0.03in}p{0.03in}p{0.24in}p{0.40in}p{0.60in}p{0.60in}p{0.40in}}

$\phi$ &$a_{5}$ & $a_{5}$ &$a_{1}$,$a_{5}$ &$a_{2}$,$a_{5}$ &$a_{1}$,$a_{2}$,$a_{4}$,$a_{5}$ & $a_{1}$,$a_{2}$,$a_{4}$,$a_{5}$ &$a_{1}$,$a_{2}$,$a_{5}$\\

&$\phi$ &$\phi$ &$a_{1}$,$a_{5}$ &$a_{2}$,$a_{5}$ &$a_{2}$,$a_{4}$,$a_{5}$ &$a_{1}$,$a_{2}$,$a_{4}$,$a_{5}$ &$a_{1}$,$a_{2}$,$a_{5}$\\

&     &$\phi$ &$a_{1}$,$a_{5}$ &$a_{2}$,$a_{5}$ &$a_{2}$,$a_{4}$,$a_{5}$ &$a_{1}$,$a_{2}$,$a_{4}$,$a_{5}$ & $a_{1}$,$a_{2}$,$a_{5}$\\

& & &$\phi$ &$a_{1}$,$a_{2}$,$a_{5}$ &$a_{2}$,$a_{4}$,$a_{5}$ &$a_{2}$,$a_{4}$,$a_{5}$ &$a_{2}$,$a_{5}$\\

& & & &$\phi$ &$a_{1}$,$a_{4}$,$a_{5}$ &$\phi$ &$a_{1}$,$a_{2}$,$a_{5}$\\

& & & & &$\phi$ &$a_{2}$,$a_{5}$ &$\phi$ \\
& & & & & &$\phi$ &$a_{4}$,$a_{5}$ \\
& & & & & & &$\phi$
\end{array}\right)$
\par\end{center}

\begin{figure}[h]
\centering
\caption {The discernibility matrix for Table \ref{externaltable}.}\label{externalmatrix}
\end {figure}

According to the approach proposed in Section \ref{roughset}, we find that $a_{5}$ is the core attribution, which indicates that the variance of instructions retired in different processes is the root cause of code region 11.

\begin{figure}
\centering
\includegraphics [width=3in]{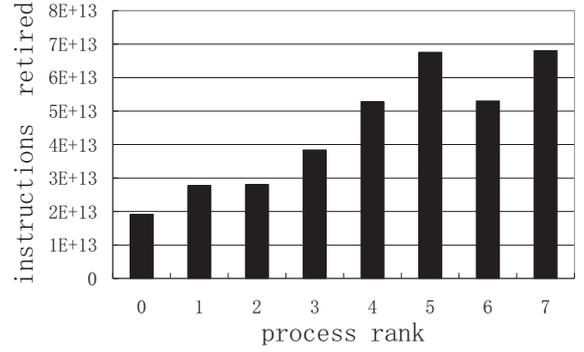}\\
\caption {The variance of instructions retired of \emph{code region 11} in different processes.}\label{instruction}
\end {figure}

Fig.\ref{instruction} verifies our analysis, from which we can discover obvious differences of instructions retired of \emph{code region 11} among different processes.

\begin{figure} [h]
\centering
\includegraphics[width=2.5in]{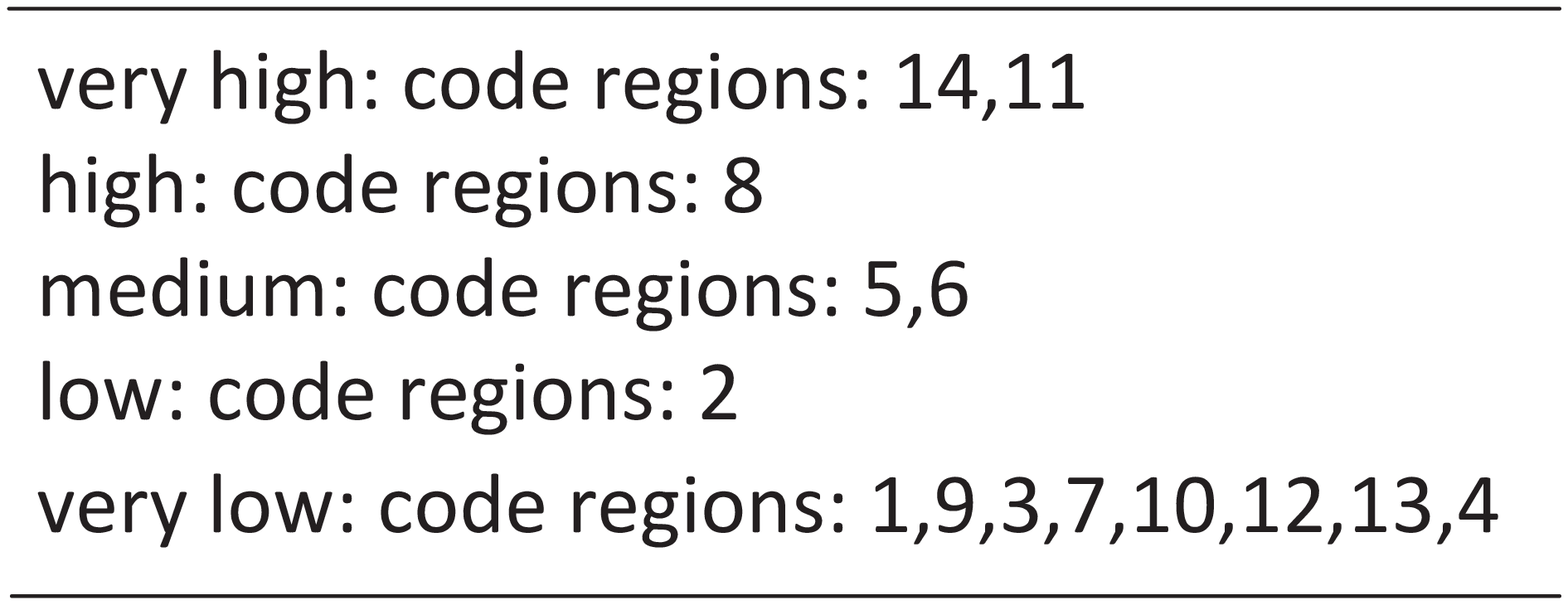}
\caption {The analysis results of the k-means clustering approach.}\label{internal_analysis}
\end {figure}

Using the K-means clustering approach,  AutoAnalyzer outputs the analysis result for each code region of ST, which is shown in Fig.\ref{internal_analysis}. The severity degree of \emph{code region 14, code region 11, code region 8} is larger than medium, respectively. According to the analysis result, we confirm that \emph{code regions 14, code region 11 and code region 8} are CCR. Since code region 11 is nested within code region 14 and the severity degree of code region 11 is the same as code region 14, so code region 11 is a CCCR. Since no code region is nested in code region 8, so code region 8 is also a CCCR. We focus on code region 8 and code region 11 for performance optimization.

\begin{figure} [h]
\centering
\includegraphics[width=3.0in]{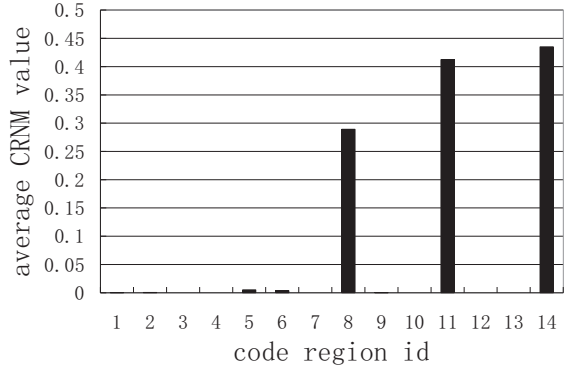}
\caption {The average CRNM of each code region.} \label{ST-CRNM}
\end {figure}

We analyze the root causes of disparity bottlenecks with the rough set approach. The decision table is shown in Table \ref{decision_internal}. In the decision table, attribution $a_{k, k=1,2,3,4,5}$ represents L1 Cache miss rate, L2 cache miss rate, disk I/O quantity, network I/O quantity, and instructions retired, respectively.

\begin{table}[h]
\centering
\caption{Decision table used for searching disparity bottlenecks.}\label{decision_internal}
\begin{tabular}{|c|c|c|c|c|c|c|}
\hline
$ID$ & $a_{1}$ & $a_{2}$ & $a_{3}$ & $a_{4}$ & $a_{5}$ & $D$\\
\hline
1 & 0 & 0 & 0 & 0 & 0 & 0\\
\hline
2 & 1 & 0 & 0 & 0 & 0 & 0\\
\hline
3 & 0 & 0 & 0 & 0 & 0 & 0\\
\hline
4 & 0 & 0 & 0 & 0 & 0 & 0\\
\hline
5 & 1 & 1 & 0 & 0 & 1 & 0\\
\hline
6 & 1 & 0 & 0 & 0 & 1 & 0\\
\hline
7 & 0 & 0 & 0 & 0 & 0 & 0\\
\hline
8 & 0 & 0 & 1 & 0 & 1 & 1\\
\hline
9 & 1 & 0 & 0 & 0 & 0 & 0\\
\hline
10 & 1 & 0 & 0 & 0 & 0 & 0\\
\hline
11 & 1 & 1 & 0 & 0 & 1 & 1\\
\hline
12 & 0 & 0 & 0 & 0 & 0 & 0\\
\hline
13 & 0 & 0 & 0 & 0 & 0 & 0\\
\hline
14 & 1 & 1 & 0 & 0 & 1 & 1\\
\hline
\end{tabular}
\end{table}

According to the approach proposed in Section \ref{roughset}, we find that $\{a_{2}, a_{3}\}$ is the core attributions, which indicates high L2 cache miss rate and high disk I/O quantity are the root causes of disparity bottlenecks. Then we search the decision table and find that the root cause of code region 8 is high disk I/O quantity and the root cause of code region 11 is high L2 cache miss. From the performance data, we can observe that the disk I/O quantity of code region 8 is as high as $106 G$ and the L2 cache miss rate of  code region 11 is high as 17.8\%.


In order to eliminate the dissimilarity bottleneck---code region 11, we replace the static load dispatching in the master process, adopted in the original program, with a dynamic load dispatching mode.
After the optimization, we use AutoAnalyzer to analyze the optimized code again. The analysis results show that all processes,
excluding the code regions in the master process responsible for the management routines, are classified into one cluster,
  indicating that all processes have the similar performance with balanced workloads.

We take the following approaches to optimizing the disparity bottlenecks---code region 8 and code region 11.  First, we improve \emph{code region 8} by buffering as many as data into the memory. Second, we improve the data locality of \emph{code region 11} by breaking the loops into small ones and rearranging the data storage.

We use AutoAnalyzer to analyze the optimized code again. The new analysis results show code region 8 is not a disparity bottleneck again, while code region 11 is still a disparity bottleneck, but the average CRNM value of code region 11 decreases from 0.41 to 0.26.  The  root cause of code region 11 is no longer the high L2 caches miss rate, but the large quantity of instructions retired.

\begin{figure} [h]
\centering
\includegraphics[width=3.5in]{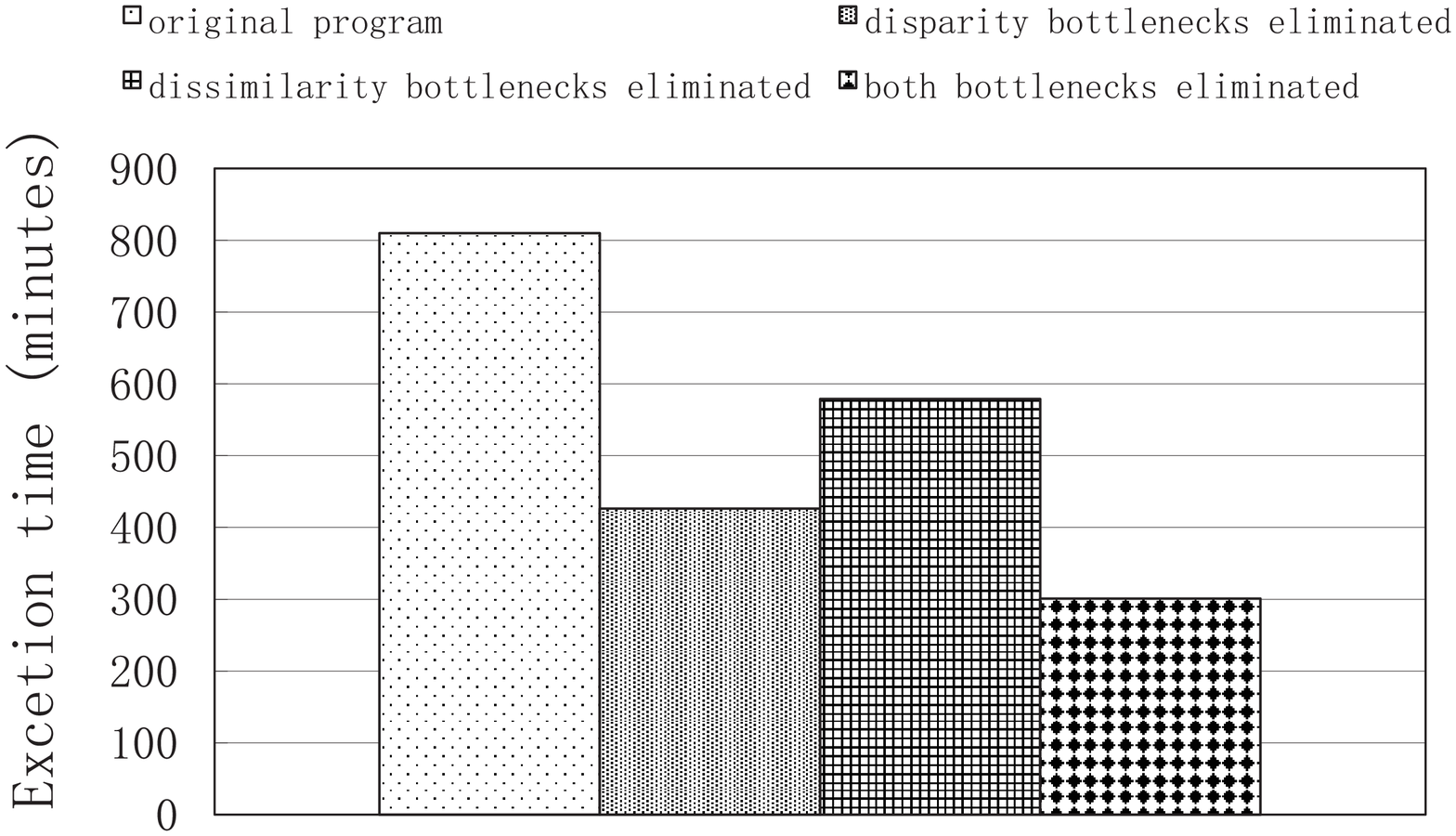}\\
\caption {ST performance before and after the optimization.}\label{ST_performance}
\end {figure}

Fig.\ref{ST_performance} shows the performance of \emph{ST} before and after the optimization. With the disparity bottlenecks eliminated, the performance of \emph{ST} rises by 90\% in comparison with the original program. With the dissimilarity bottlenecks eliminated, the performance of ST rises by 40\% in comparison with the original program. With both disparity and dissimilarity bottlenecks eliminated, the performance of ST rise by 170\% in comparison with the original program.

\subsubsection{A case study of ST with fine-grain code regions }\label{granularity}

\begin{figure} [h]
\centering
\includegraphics[width=3.0in]{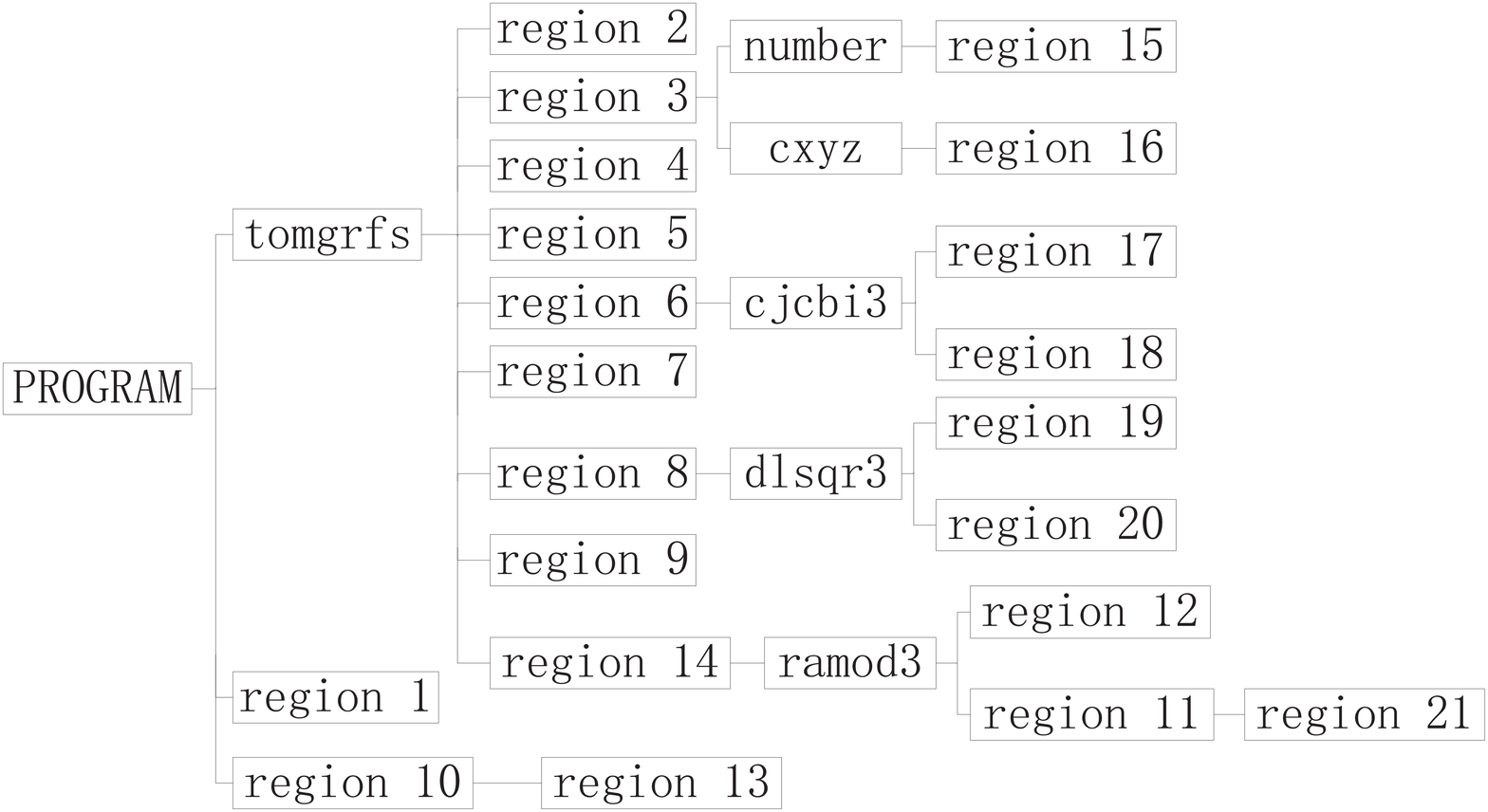}\\
\caption {The refined code region tree.}\label{ST_tree_refined}
\end {figure}

In this subsection, on a basis of the code region tree shown in Fig.\ref{STtree}, we divide the program  into fine-grained code regions, which is shown in Fig. \ref{ST_tree_refined}. For saving time, we choose the shot number as 300, and the run time of application is about 9815.52454 seconds.
Please note that with the exception of newly added code regions, the same code regions in Fig.\ref{STtree} and Fig. \ref{ST_tree_refined} keep the same ID.

We use the simplified OPTICS clustering algorithm to find dissimilarity bottlenecks. From the analysis result, we can find that code region 14, code region 11, and code region 21 are CCRs. Since code region 21 is nested within code region 11 and the latter is also nested within code region 14, we confirm that code region 21 is a CCCR, which is the location of the problem.

From  Fig.\ref{STtree} and Fig. \ref{ST_tree_refined}, we can observe the newly identified dissimilarity bottleneck--- code region 21 is nested within code region 11, which is identified as a dissimilarity bottlenecks in Section \ref{ST_B} when a  coarse-grain code region tree is adopted.


We also use the k-means clustering approach to locateing disparity bottlenecks. From the analysis results, we conclude code region 19 and code region 21 are disparity bottlenecks.

\begin{figure}
\centering
\includegraphics [width=3.0in]{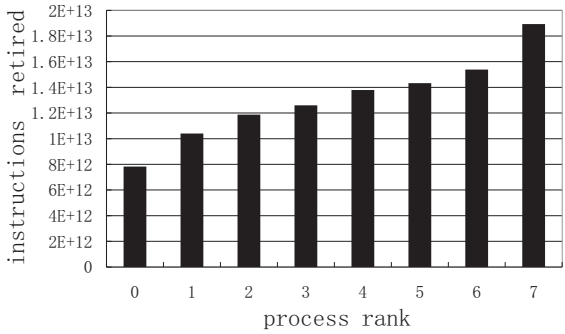}\\
\caption {The variance of instructions retired of \emph{code region 21} in different processes.}\label{21_instruction}
\end {figure}

From  Fig.\ref{STtree} and Fig. \ref{ST_tree_refined}, we can observe the newly identified disparity bottlenecks---code region 19 and code region 21 are nested within code region 8 and code region 14, respectively, which are identified as disparity  bottlenecks in Section \ref{ST_B} when a  coarse-grain code region tree is adopted.   These results show our two-round analysis, introduced in Section \ref{implementation}, indeed can refine the scope of both dissimilarity bottlenecks and disparity bottlenecks. Fig.\ref{21_instruction} shows the variance of instructions retired of code region 21 in different processes.

\subsection{NPAR1WAY} \label{NPAT1WAY}
NPAR1WAY is a module of the SAS (Statistical Analysis System) responsible for reading, writing, managing, analyzing, and displaying data. SAS is Widely used in data and statistical analysis. The parallel NPAR1WAY module uses MPI to calculate the exact p-value to achieve high performance. AutoAnalyzer divides the whole program into 12 code regions to separate functions, subroutines, and outer loops.

Out testbed is a small-scale cluster system. Each node has two processors, each of which is a 2 GHz Intel Xeon Processor E5335 with quad cores, $128 KB$ L1 data cache, $128 KB$ L1 instruction cache, and 8 MB L2 cache. The operating system is Linux 2.6.19.

\subsubsection{Bottleneck detection} \label{NPAR1Way_bottlenecks}

The analysis results of AutoAnalyzer shows all processes are classified into one cluster, which indicates that no dissimilarity bottleneck exists. AutoAnalyzer also analyzes the application performance from the perspective of each code region.
   The analysis results show that the severity degrees of code region 3 and code region 12 are larger than medium, and we consider them as CCR. Because there are no nested code regions in code region 3 and code region 12, both of two code regions are CCCRs, which we consider disparity bottlenecks.

\begin{figure} [h]
\centering
\includegraphics[width=3in]{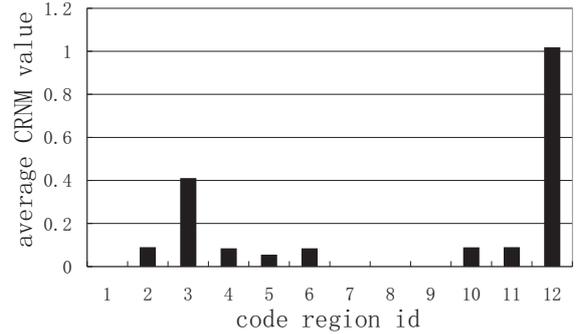}\\
\caption {The average CRNM of each code region in eight processes.}\label{NPAR1WAY_CNRM}
\end {figure}

We also use the rough set approach to uncover the root causes of disparity bottlenecks. In the decision table, the attributes $a_{k, k=1,2,3,4,5}$  represents L1 cache miss rate, L2 cache miss rate, disk I/O quantity, network I/O quantity, and instructions retired, respectively.

Through analyzing the discernibility matrix, we conclude that $\{a_{4}, a_{5}\}$ are the core attributions, which indicates that both high network I/O quantity and high instructions retired are root causes of the disparity bottlenecks. Then we search the decision table and find that code region 3 has high quantity of instructions retired. Meanwhile code region 12 has both high quantity of instructions retired and high network I/O quantity. From the performance data, we can see that instructions retired of code region 3 and code region 12 take up 26\% and 60\% of the total instructions retired of the program, respectively. At the same time, the network I/O quantity of code region 12 takes up  70\% of the total network I/O quantity of the program.


\subsubsection{The performance optimization}

According to the root causes uncovered by AutoAnalyzer, we optimize the code to eliminate the disparity bottlenecks. The performance of NPAR1WAY rises by 20\% after the optimization.

We optimize code region 3 and code region 12 by eliminating \emph{redundant common expressions}. For example, there is one common multiply expression occurring three times in code region 3. We use one variable to store the results of the multiply expression at its first appearance, and later directly use the variable to avoid subsequent redundant computation. In this way, we can decrease massive instructions by eliminating redundant common expressions in deep loops.

Then we analyze the code again. For the optimized code region 3, the analysis results show that the quantity of instructions retired and the wall clock time are reduced by 36.32\% and 20.33\%, respectively. For the optimized code region 12, the analysis results show that the instructions retired and the wall clock time are reduced by 16.93\% and 8.46\%, respectively. For code region 12, we fail to eliminate high network I/O quantity.


\subsection{Analysis of an open source application---MPIBZIP2} \label{MPIBZIP2}
\begin{figure} [h]
\centering
\includegraphics[width=2.5in]{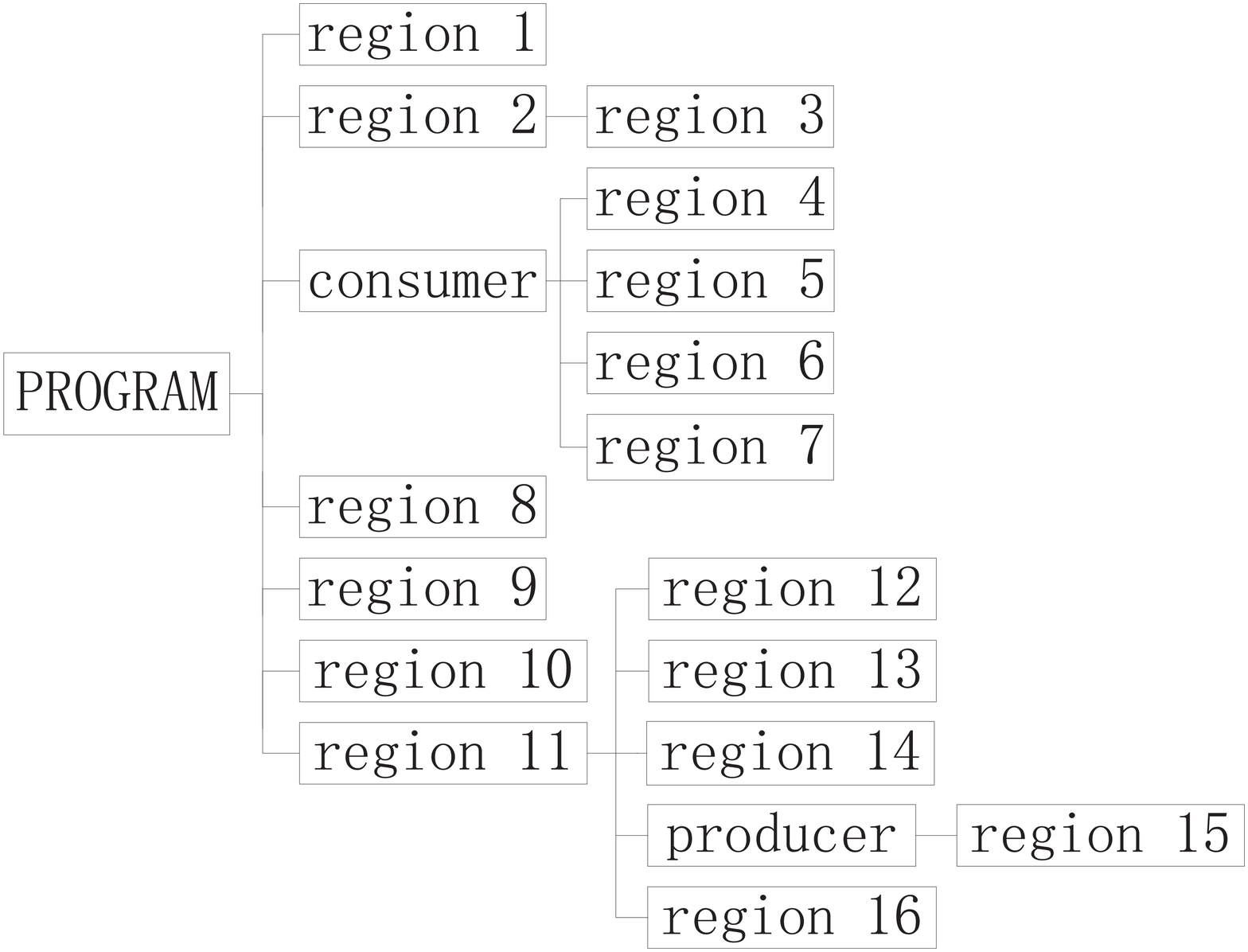}\\
\caption {The code region tree of TMPIBzip2.}\label{MPIBzip2_tree}
\end {figure}

MPIBZIP2 is a parallel implementation of the bzip2 block-sorting file compressor that uses MPI and achieves significant speedup on cluster machines. The output is fully compatible with the regular bzip2 data so any files created with MPIBZIP2 can be uncompressed by bzip2 and vice-versa. This software is open source  and distributed under a BSD-style license.
AutoAnalyzer divides the whole program into 16 code regions to separate functions, subroutines, and outer loops. Fig.\ref{MPIBzip2_tree} shows the code region tree. Out testbed is just the same as that in Section \ref{NPAT1WAY}.

\begin{figure} [h]
\centering
\includegraphics[width=3.0in]{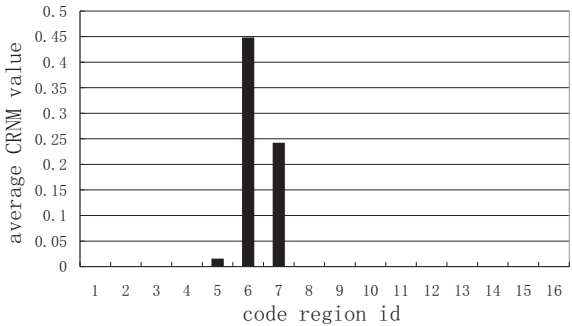}\\
\caption {The average CRNM of each code region of MPIBZIP2.}\label{MPIBZIP2_CRNM}
\end {figure}

Excluding the code regions that are responsible for management routines in the master process,  we use the simplified OPTICS clustering algorithm to find dissimilarity bottlenecks. From the analysis result, we find all processes are classified into one cluster, and we confirm that there are no dissimilarity bottlenecks in MPIBZIP2. We also use the K-means clustering approach to analyzing the disparity bottlenecks. The analysis results show that the severity degrees of code region 6, and code region 7 are larger than medium, and we consider them as CCR. Since there are no nested code regions in code region 6 and code region 7, both of two code regions are CCCRs, which we consider disparity bottlenecks. Fig.\ref{MPIBZIP2_CRNM} shows the average CRNM of each code region of MPIBZIP2.

We uncover the root causes of disparity bottlenecks with the rough set approach.
In the decision table, the attributes $a_{k,k=1,2,3,4,5}$ represents L1 cache miss rate, L2 cache miss rate, disk I/O quantity, network I/O quantity and instructions retired, respectively.
Through analyzing the discernibility matrix, we conclude that $\{a4,a5\}$ are the core attributions, which indicates that network I/O quantity and instructions retired are root causes of the disparity bottlenecks.
Then we search the decision table and find that the root cause of code region 6 is high quantity of instructions retired and the root cause of code region 7 is high network I/O quantity.
From the performance data, we also observe that instructions retired of code region 6 take up 96\% of the total instructions retired of the program.
At the same time,the network I/O quantity of code region 7 take up 50\% of the total network I/O quantity of the program.

Through reading the source code, we found out that code region 6 calls the $BZ2\_bzBuffToBuffCompress()$ function to compress the data. $BZ2\_bzBuffToBuffCompress()$ is a third-party function and packaged in the static library libbz2.a of bzip2. Code region 7 call $MPI\_Send()$ to send the compressed data to the master process. Those two bottlenecks are difficult to optimize. For the first bottleneck, we need to improve the mature compression algorithm;  for the second bottleneck, we need to decrease the data transferred to the master process, however the data has been compressed. We fail to optimize the code.

\subsection{Effect of different metrics on bottleneck detections} \label{different_metrics}

For three applications, we investigate the effect of different metrics on locating bottlenecks.
For ST, NPAR1WAY, and MPIBZIP2, the number of code regions is 14, 12, and 16, respectively.
For ST, we perform the experiments on the same testbed as that in Section \ref{ST}, but the shot number is changed from 627 to 300 for saving time. For two other applications, the testbed is the same as that in Section \ref{NPAT1WAY}.

We choose the CRNM value, the CPI, and the wall clock time of each code region as the main performance measurement to locate disparity bottlenecks,  respectively.

\begin{figure} [h]
\centering
\includegraphics[width=3.0in]{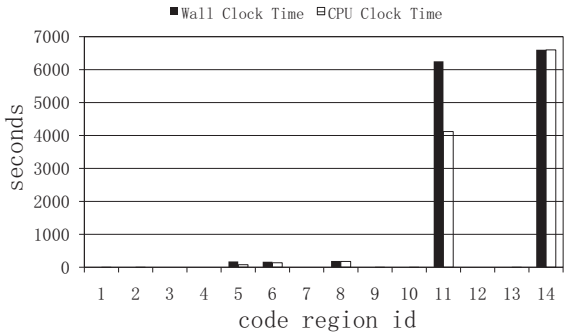}\\
\caption {The average wall clock time and CPU clock time of each code region of ST.}\label{ST_300_wall_clock}
\end {figure}

Our experiment shows CRNM is more valuable than CPI or the wall clock time on locating disparity bottlenecks. For example, for ST, using CRNM, AutoAnalyzer identifies code region 8, code region 11, and code region 14 as CCR, and we significantly improve the application performance through optimizing them, as shown in Section \ref{ST_B}; using the average wall clock time of each code region, AutoAnalyzer identifies code region 2,5, 6, 10 as disparity bottlenecks in addition to code region 8, 11 and 14. From Fig. \ref{ST_300_wall_clock}, we can observe code region 2, 5, 6, 10 take up trivial proportion of the running time of the application. Using CPI, AutoAnalyzer identifies code region 2, 8 as disparity bottlenecks, while code region 11 and code region 14, which take up most of the running time of the application, are ignored.

\begin{figure} [h]
\centering
\includegraphics[width=3.0in]{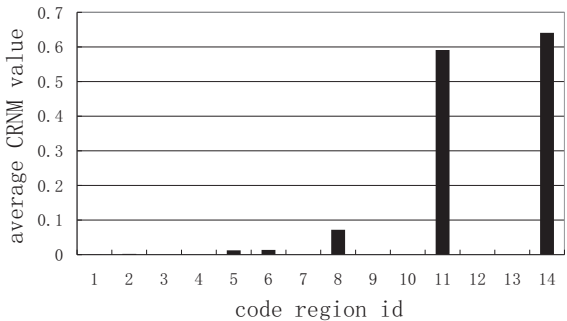}\\
\caption {The average CRNM of each code region of ST.}\label{CRNM}
\end {figure}

\begin{figure} [h]
\centering
\includegraphics[width=3.0in]{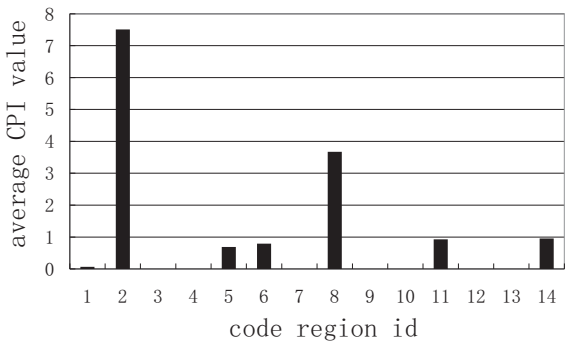}\\
\caption {The average CPI of each code region of ST.}\label{CPI}
\end {figure}

Fig.\ref{ST_300_wall_clock}, Fig.\ref{CRNM}, and Fig.\ref{CPI} show the average wall clock time and CPU clock time, the average CRNM,  and CPI of each code region of ST, respectively.

CRNM is more valuable than CPI or wall clock time on locating disparity bottlenecks because of the following two reasons: first, by using the ratio of \emph{the wall clock time of a code region} to \emph{the wall clock time of the whole program}, our metrics can judge the performance contribution of a code region to the overall performance of a program. Second, CPI measures the efficiency of instruction execution. Derived from the total instructions retired and the total executing cycles, CPI is a basic metric that reflects all hardware events: cache or TLB miss, cache line invention, pipeline stall caused by data dependency or branches misprediction and so on. So our normalized CPI represents a measurement of the importance of a code region to the overall performance of the application.

\begin{figure} [h]
\centering
\includegraphics[width=3.0in]{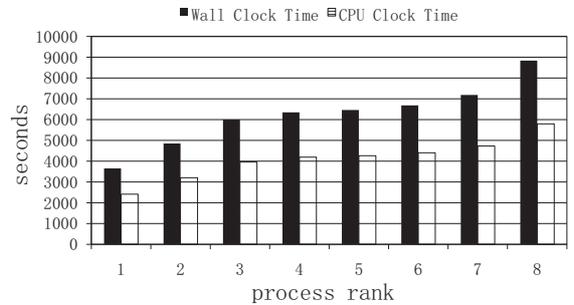}\\
\caption {The wall clock time and CPU clock time of code region 11 of ST in different processes.}\label{wall_CPU_code_region_11}
\end {figure}

We choose the wall clock time and the CPU clock time as the main measurement to locate dissimilarity bottlenecks,  respectively. For three applications, we utilize two metrics to locate dissimilarity bottlenecks, respectively.  As an example, Fig.\ref{ST_300_wall_clock} compares the average wall clock time and the average CPU clock time of each code region of ST, and Fig.\ref{wall_CPU_code_region_11} shows the wall clock time and the CPU clock time of code region 11 of ST in different processes, which is identified as a dissimilarity bottleneck in Section \ref{ST_B}.  Though two measurements have some differences,  our results show they have the same effects on locating dissimilarity bottlenecks.

\section{Conclusions} \label{conclusion}
This paper presented a series of innovative methods in automatic performance debugging of SPMD-style parallel programs.
For SPMD-style parallel applications,  we utilized two effective clustering algorithms to investigate the existence of two types of bottlenecks: dissimilarity bottlenecks that cause process behavior dissimilarity and disparity bottlenecks that cause code region behavior disparity; if there are bottlenecks, we presented two searching algorithms to locate performance bottlenecks. On a basis of the rough set theory, we proposed an innovative approach to automatically uncovering root causes of bottlenecks. We designed and implemented AutoAnalyzer. On the cluster systems with two different configurations, we used two production applications and one open source code---MPIBZIP2 to verify the effectiveness and correctness of our methods. Meanwhile, we also investigate the effects of different metrics on locating bottlenecks, and our experiment results showed for three applications, our proposed metrics---CRNM outperforms CPI and wall clock time in terms of locating disparity bottlenecks; the wall clock time and the CPU clock time have the same effects on locating dissimilarity bottlenecks.

     In the near future, we will extend our method to more generalized parallel applications beyond the SPMD style.


\section*{Acknowledgment}

We are very grateful to anonymous JPDC reviewers for their constructive comments. This work is supported by the NSFC projects(Grant No.60703020 and Grant No.60933003), the Chinese national 973 project(Grant No.2011CB302500), and the Chinese national 863 project(Grant No.2009AA01Z128).

\


\section*{References}
\begin{thebibliography} {99}
\bibitem{ref_1}
M. Ankerst, M. M. Breunig, H. Kriegel, and J. Sander, OPTICS: ordering points to identify the clustering structure. SIGMOD Rec. 28, 2 (Jun. 1999) 49-60.
\bibitem{ref_2}
B. Mohr and F. Wolf. KOJAK, A Tool Set for Automatic Performance Analysis of Parallel Applications. In Ninth Intl. Euro-Par Conference (Euro-Par 2003), Klagenfurt, Austria, August 2003.
\bibitem{ref_3}
F. Wolf,  and B. Mohr, Automatic performance analysis of hybrid MPI/OpenMP applications. J. Syst. Archit. 49, 10-11 (Nov. 2003) 421-439.
\bibitem{ref_4}
 T. Fahringer, M. Gerndt, B. Mohr, F. Wolf, G. Riley, and J. L. Traff. Knowledge specification for automatic performance analysis: APART technical report, revised edition. Tech. Rep. FZJ-ZAM-IB-2001-08, Forschungszentrum J¡§ulich GmbH, Aug. 2001.
\bibitem{ref_5}
F. Wolf,  B. Mohr, J. Dongarra, and S. Moore,  Automatic analysis of inefficiency patterns in parallel applications: Research Articles. Concurr. Comput. : Pract. Exper. 19, 11 (Aug. 2007) 1481-1496.
\bibitem{ref_6}	
B. Mohr. OPARI-OpenMP Pragma and Region Instrumentor. Available from < http://www.fz-juelich.de/jsc/kojak/opari/>.
\bibitem{ref_7}	
J. K. Hollingsworth,  and B. P. Miller, Dynamic control of performance monitoring on large scale parallel systems. In Proceedings of ICS '93 (July.1993) 185-194.
\bibitem{ref_8}
K. L. Karavanic,  and B. P. Miller, Improving online performance diagnosis by the use of historical performance data. In Proceedings of SC 09 (Nov. 1999), 42.
\bibitem{ref_9}	
H. W. Cain, B. P. Miller, and B. J.  Wylie, A Callgraph-Based Search Strategy for Automated Performance Diagnosis. In Proceedings of ICPP 2000 (August 29 - September 01, 2000) 108-122.
\bibitem{ref_10}
A. R. Bernat and B. P. Miller, Incremental call-path profiling. In Technical report, University of Wisconsin, 2004.
\bibitem{ref_11}
P. C. Roth, and B. P. Miller, Deep Start: A Hybrid Strategy for Automated Performance Problem Searches. In Proceedings of the 8th Euro-Par (Aug. 2002) 86-96.
\bibitem{ref_12}	
J.A. Hartigan, and M.A.  Wong, A k-means clustering algorithm. In  Applied Statistics, 28 (1979) 100-108.
\bibitem{ref_14}	
J. Mellor-Crummey, R. J. Fowler, G. Marin, and  N. Tallent, HPCVIEW: A Tool for Top-down Analysis of Node Performance. J. Supercomput. 23, 1 (Aug. 2002) 81-104.
\bibitem{ref_15}	J. Komorowski, Z. Pawlak, L. Polkowsk and A. Skowron, Rough sets: A tutorial. Springer-Verlay (1999) 3-9.

\bibitem{ref_16}	M. Calzarossa, L. Massari, and D. Tessera, A methodology towards automatic performance analysis of parallel applications. Parallel Comput. 30, 2 (Feb. 2004) 211-223.

\bibitem{ref_17}	N. R. Tallent,  and J. M. Mellor-Crummey,  Effective performance measurement and analysis of multithreaded applications. In Proceedings of the 14th PPoPP (Feb.2009) 229-240.

\bibitem{ref_19}	Z. Pawlak, Rough sets. In International Journal of Information and Computer Science, 11(1982) 341-356.
\bibitem{ref_21} V.V. Dimakopoulos, E. Leontiadis, and G. Tzoumas, A Portable C Compiler for OpenMP V.2.0. In Proc. of the 5th European Workshop on OpenMP (EWOMP03), Aachen, Germany (October 2003).
\bibitem{ref_22} K. A. Huck,  and A. D. Malony, PerfExplorer: A Performance Data Mining Framework For Large-Scale Parallel Computing. In Proceedings of SC 05(Nov. 2005), 41.
\bibitem{ref_23} K. A. Huck,  O. Hernandez, and V. Bui, S. Chandrasekaran, B. Chapman,  A. D. Malony, L. C. McInnes, and B. Norris, Capturing performance knowledge for automated analysis. In Proceedings of SC08 (Nov.2008), NJ 1-10.
\bibitem{ref_24} K. A. Huck, A. D. Malony, S. Shende, and A. Morris, Knowledge support and automation for performance analysis with PerfExplorer 2.0. Sci. Program. 16, 2-3 (Apr. 2008) 123-134.
\bibitem{ref_25}	S. Moore, F. Wolf, J. Dongarra, S. Shende, A. Malony, and B. Mohr. A Scalable Approach to MPI Application Performance Analysis. In LNCS, 3666 (2005) 309-316.
\bibitem{ref_26}  B. Di Martino,  E. Mancini,  M. Rak, R. Torella, and U. Villano, Cluster systems and simulation: from benchmarking to off-line performance prediction: Research Articles. Concurr. Comput. : Pract. Exper. 19, 11 (Aug. 2007) 1549-1562.
\bibitem{ref_27} D. Rodr¨ªguez,  A Statistical Approach for the Analysis of the Relation Between Low-Level Performance Information, the Code, and the Environment. In Proceedings of the 2002 international Conference on Parallel Processing Workshops (August, 2002) 282.

\bibitem{SC_workshop}
X. Liu, J. Zhan, D. Meng, M. Zou, B. Tu, Similarity Analysis in Automatic Performance Debugging of SPMD Parallel Programs.
Workshop on Node Level Parallelism for Large Scale Supercomputers, Co-located with ACM/IEEE SC08.

\bibitem{ref_28} J. Vetter,  Performance analysis of distributed applications using automatic classification of communication inefficiencies. In Proceedings of the 14th ICS (May. 2000) 245-254.

\bibitem{preciseTracer}  Z. Zhang, J. Zhan, Y. Li, L. Wang, D. Meng, B. Sang, Precise request tracing and performance debugging for multi-tier services of black boxes. In Proceedings of the 39th DSN (June 2009) 337-346

\bibitem{ref_29} D. H.  Ahn, and  J. S.  Vetter, Scalable analysis techniques for microprocessor performance counter metrics. In Proceedings of SC 02 1-16.
\bibitem{ref_30}	H.-L. Truong and T. Fahringer. SCALEA: a Performance Analysis Tool for Parallel Programs. In Concurrency and Computation: Practice and Experience, 15(11-12) (2003) 1001-1025.
\bibitem{ref_31} H.-L. Truong, T. Fahringer, Soft Computing Approach to Performance Analysis of Parallel and Distributed Programs. In proceedings of Euro-Par 2005 50-60
\bibitem{ref_32} T. Sherwood, E. Perelman, G. Hamerly, and B. Calder,Automatically characterizing large scale program behavior. In Proceedings of the 10th ASPLOS (Oct. 2002) 45-57.
\bibitem{ref_33} J.K. Hollingsworth, M. Steele, Grindstone: A test suite for parallel performance tools, Computer Science Technical Report CS-TR-3703, University of Maryland, October 1996.
\bibitem{ref_34} T. Fahringer, M. Geissler, G. Madsen, H. Moritsch and C. Seragiotto. On using Aksum for semi automatically searching of performance problems in parallel and distributed programs. In Proceedings of 11th PDP (2003) 385-392.

\bibitem{debug_mapreduce} S. Babu,  Towards automatic optimization of MapReduce programs. In Proceedings of 1st SoCC (2010) 137-142.

\bibitem{ref_35} A. Tiwari, C. Chen, J. Chame, M. Hall, J. K. Hollingsworth,  A Scalable Autotuning Framework for Compiler Optimization ,IPDPS 2009 (May. 2009).
\bibitem{ref_36} 	S. S. Shende,  and A. D. Malony, The TAU parallel performance system. In The International Journal of High Performance Computing Applications, 20, 2 (2006) 287-311,.
\bibitem{ref_37} K. A. Huck, and A. D. Malony. Performance forensics: knowledge support for parallel performance data mining. http://ix.cs.uoregon.edu/~khuck/papers/parco2007.pdf.
\bibitem{ref_38} L. Li, and A. D. Malony,  Knowledge engineering for automatic parallel performance diagnosis: Research Articles. Concurr. Comput. : Pract. Exper. 19, 11 (Aug. 2007) 1497-1515.
\bibitem{ref_39} L. Li and A. D. Malony. Automatic Performance Diagnosis of Parallel Computations with Compositional Models. In Proc. IPDPS 07 (2007).
 \bibitem{hpctookit}
    L. Adhianto, S. Banerjee, M. Fagan, M. Krentel, G. Marin, J. Mellor-Crummey, and N. R. Tallent, HPCTOOLKIT: tools for performance analysis of optimized parallel programs. Concurr. Comput. : Pract. Exper. 22, 6 (April 2010) 685-701.

    \bibitem{JSC}
B. Tu, J. Fan, J. Zhan, and X. Zhao, Performance analysis and optimization of MPI collective operations on multi-core clusters, The Journal of Supercomputing, DOI: 10.1007/s11227-009-0296-3, online.

\bibitem{tu_pdp}
B. Tu, J. Fan, J. Zhan, and X. Zhao, Accurate Analytical Models for Message Passing on Multi-core Clusters, In Proceedings of the 17th PDP (Feb 2009) 133-139.

\bibitem{SPMD}
F. Darema, SPMD model: Past, present and future, Recent Advances in Parallel Virtual
Machine and Message Passing Interface: Eighth European PVM/MPI Users' Group Meeting,
Santorini/Thera, Greece, 2001.

\bibitem{Granules}
S. Pallickara, J. Ekanayake, and G. Fox, Granules: A Lightweight Runtime for Scalable Computing with Support for Map-Reduce, Cloud Computing and Software Services: Theory and Techniques: CRC Press (Taylor and Francis), 07/2010.

\bibitem{INTP}
J. Ekanayake, S. Pallickara, and G.C. Fox, Performance of Data Intensive Supercomputing Runtime Environments , Bloomington, IN, Indiana University, 08/01/2008.

\bibitem{Transformer}
P. Wang, D. Meng, J. Han, J. Zhan, B. Tu, X. Shi, and L. Wan, Transformer: A New Paradigm for Building Data-Parallel Programming Models. IEEE Micro 30, 4 (July 2010) 55-64.

\bibitem{SC-MTC}
L. Wang, J. Zhan, W. Shi, Y. Liang, and L. Yuan, In cloud, do MTC or HTC service providers benefit from the economies of scale?. In Proceedings of the 2nd MTAGS (2009). 10 pages.


\bibitem{DawningCloud}
L. Wang, J. Zhan, W. Shi, and Y. Liang, In Cloud, Can Scientific Communities Benefit
from the Economies of Scale? Accepted by IEEE Transaction on Parallel and Distributed Systems. March, 2011.

\bibitem{Mapreduce}
J. Dean and S. Ghemawat, Mapreduce: Simplified data processing on large clusters,
Communications of the ACM, 51 (January 2008) 107-113.
\bibitem{core}
X. Hu, N. Cercone, Learning in Relational Databases: a Rough Set Approach,
Computational Intelligence, 2(1995) 323-337


\bibitem{CoRR}
X. Liu, Y. Lin, J. Zhan, B. Tu, D. Meng, Automatic Performance Debugging of SPMD Parallel Programs. Technical Report, Institute of Computing Technology, Chinese Academy of Sciences. http://arxiv.org/abs/1002.4264





\end{thebibliography}
\end{document}